\documentstyle[12pt,epsfig]{article}
\newcommand{\be}{\begin{equation}}
\newcommand{\ee}{\end{equation}}
\newcommand{\bea}{\begin{eqnarray}}
\newcommand{\eea}{\end{eqnarray}}

\newcommand{\beq}{\begin{equation}}
\newcommand{\eeq}{\end{equation}}

\newcommand{\nn}{\nonumber}

\def\fun#1#2{\lower3.6pt\vbox{\baselineskip0pt\lineskip.9pt
\ialign{$\mathsurround=0pt#1\hfil##\hfil$\crcr#2\crcr\sim\crcr}}}

\begin{document}

\title{Quark--antiquark
states and their radiative transitions in terms of the spectral
integral equation.\\ {\Huge II.} Charmonia.}
\author{V.V. Anisovich, L.G. Dakhno, M.A. Matveev,\\ V.A. Nikonov
and A.V. Sarantsev}

\date{\today}
\maketitle

\begin{abstract}
In the precedent paper of the authors (hep-ph/0510410), the
$b\bar b$ states were treated in the framework of the spectral
integral equation, together with simultaneous
calculations of radiative decays of the considered bottomonia.
In the present paper, such a study is carried out for the
charmonium $(c\bar c)$ states. We reconstruct the interaction in the
$c\bar c$-sector on the basis of data for the charmonium levels with
$J^{PC}=0^{-+}$, $1^{--}$, $0^{++}$, $1^{++}$, $2^{++}$,
$1^{+-}$ and radiative transitions
$\psi(2S)\to\gamma\chi_{c0}(1P)$, $\gamma\chi_{c1}(1P)$,
$\gamma\chi_{c2}(1P)$, $\gamma\eta_{c}(1S)$ and
$\chi_{c0}(1P)$, $\chi_{c1}(1P)$, $\chi_{c2}(1P)\to\gamma J/\psi$.
The $c\bar c$ levels and their wave functions  are calculated for the
radial excitations with $n\le 6$. Also, we determine the $c\bar c$
component
of the photon wave function using the $e^+e^-$ annihilation data:
$e^+e^- \to J/\psi(3097)$, $\psi(3686)$, $\psi(3770)$, $\psi(4040)$,
$ \psi(4160)$, $\psi(4415)$ and perform the calculations of the partial
widths of the two-photon decays for the $n=1$ states:
$\eta_{c0}(1S)$, $\chi_{c0}(1P)$, $\chi_{c2}(1P)\to\gamma\gamma$, and
$n=2$ states: $\eta_{c0}(2S)\to\gamma\gamma$, $\chi_{c0}(2P)$,
$\chi_{c2}(2P)\to \gamma\gamma$.
We discuss the status of the recently observed  $c\bar c$ states
$X(3872)$ and $Y(3941)$: according to our results, the $X(3872)$ can be
either $\chi_{c1}(2P)$ or $\eta_{c2}(1D)$,  while $Y(3941)$
is  $\chi_{c2}(2P)$.
\end{abstract}

\section{Introduction}

In this paper, we continue the study initiated in
\cite{bb} for the $b\bar b$ states in the framework of the
spectral integral
equation. Here the results for the $c\bar c$-states are
presented.

In \cite{BS}, the program has been formulated for the
reconstruction of the quark--antiquark interaction based
on our knowledge of meson levels and their radiative decays.
Within this program, as a first step we have considered the bottomonia
\cite{bb}. Now we present analogous results for the charmonia. In the
subsequent publication, we suggest to give corresponding results for
 the light quark--antiquark systems.

Our study is carried out in terms of the
spectral integral technique. The application of this technique to the
composite quark--antiquark systems and its relation to  the
dispersion $N/D$-method have been discussed  in
\cite{bb,BS} --- one may find there necessary
details.

Still, let us point once again to particular properties of our approach.
 The quark--antiquark interaction given, the
spectral integral equations provides us unambiguously
with both levels and wave functions of composite systems.
But if the interaction is unknown, to reconstruct it one needs to know
the levels as well as their wave functions. Our knowledge of the
interaction of constituent quarks (in particular, long-range
interaction) is rather fragmentary, so actually the description of the
composite quark--antiquark systems means the reconstruction of
quark interaction. To this aim, one needs the information on the wave
function of a level, and the radiative decays are precisely the source
of information for the wave functions. Because of that, in our
investigation of the quark--antiquark states we rely equally upon our
knowledge of levels and radiative decays.

The method of calculation of the radiative transition amplitudes in
terms of the double dispersion integrals was developed in a number of
papers
\cite{deut,physrev,epja}, where the
important point was the
representation of the transition amplitude in the form convenient for
simultaneous fitting to the spectral integral equation
 --- it was done in \cite{YFscalar,YFtensor}.

An important information on the
$c\bar c$-meson wave function is hidden in the two-photon meson
decays: $c\bar c-meson\to \gamma\gamma$. For the calculation of
such a type processes,  the quark wave function of
the photon is needed; correspondingly, the method of reconstruction of
the $\gamma\to c\bar c$ vertices was developed in \cite{YF-g}.
The results of  calculations of the $c\bar c$-mesons are presented in
Section 2. The known levels with $J^{PC}=0^{-+}$, $1^{--}$, $0^{++}$,
$1^{++}$, $2^{++}$, $1^{+-}$ have been included in the fitting
procedure, together with the widths of radiative transitions as
follows: $\psi(2S)\to\gamma\chi_{c0}(1P)$, $\gamma\chi_{c1}(1P)$,
$\gamma\chi_{c2}(1P)$, $\gamma\eta_{c}(1S)$ and
$\chi_{c0}(1P)$, $\chi_{c1}(1P)$, $\chi_{c2}(1P)\to\gamma J/\psi$.
The masses of the states have been calculated for the radial quantum
numbers $n\le 6$ and radiative transition amplitudes for the states with
$n\le 2$. Let us emphasize that  we face rather significant
relativistic effects in the case of excited states.

The recently observed states $X(3872)$ and $Y(3941)$
\cite{Choi,Acosta,Abazov,Aubert} are lively discussed at present time
\cite{5,6,7,8,9,10,11,12,13,14}. Our calculations argue that
$X(3872)$ is either the excited $1^{++}$-state
$\chi_{c1}(2P)$, or basic $2^{-+}$-state
$\eta_{c2}(1D)$. The charmonium $Y(3941)$ state can be the
radial excited $2^{++}$ state, $\chi_{c2}(2P)$.

 We determine the $c\bar c$ component
of the photon wave function using the transitions
$e^+e^- \to J/\psi(3097)$, $\psi(3686)$, $\psi(3770)$, $\psi(4040)$,
$\psi(4160)$, $\psi(4415)$ that allow us to calculate partial widths of
the two-photon decays. We present the results for the  $1S$,
$1P$ states: $\eta_{c0}(2979)$, $\chi_{c0}(3415)$, $\chi_{c2}(3556)\to
\gamma\gamma$, and $2S$, $2P$ states: $\eta_{c0}(3594)$,
$\chi_{c0}(3849)$, $\chi_{c2}(3950)\to \gamma\gamma$.
The predictions are also given for the two-photon decays
$\eta_{c0},\chi_{c0}$, $\chi_{c2}\to  \gamma\gamma$ of the states below
$4$ GeV.

In Conclusion, we briefly summarize the results.

\section{Charmonium states}

We calculate the $c\bar c$ levels and their wave functions using two
types of the $t$-channel exchanges:  scalar and vector states,
($I\otimes I$) and ($\gamma_\mu\otimes\gamma_\mu$).  The calculations
of the $c\bar c$-systems have been carried out similarly to the
consideration of bottomonia  \cite{bb},
 with both retardation interaction
(Solution $R(c\bar c)$) and three variants of instantaneous interactions
(Solutions $I(c\bar c)$, $II(c\bar c)$ and $U(c\bar c)$). In the case of
instantaneous interactions, the $t$-channel
exchanges may be represented by using the
potentials. In the fitting to the $c\bar c$ spectra, we have applied
scalar and vector potentials of the type:
 \bea
a+b\,r+c\,e^{-\mu_c  r}+\frac{d}{r}\,e^{-\mu_d r} \ .
\label{cs-1}
\eea
The presentation of
these interactions in the momentum space, also with
 retardation effects, is given in \cite{bb}.

 The interaction parameters obtained in the fit are as follows (in
GeV units):
\be
\begin{tabular}{ccccccccc}
Interaction & Solution & \hspace{-0.5cm} $a$
& \hspace{-0.5cm} $b$ & \hspace{-0.5cm} $c$ & \hspace{-0.5cm} $\mu_c$
& \hspace{-0.5cm} $d$ & \hspace{-0.5cm} $\mu_d$ \\
\hline
$({\rm I} \otimes {\rm I})$
& \hspace{-0.5cm}
\begin{tabular}{c}
 $I(c\bar c)$ \\ $II(c\bar c)$ \\ $U(c\bar c)$ \\ $R(c\bar c)$
\end{tabular}
& \hspace{-0.5cm}
\begin{tabular}{c}
                    -1.980 \\ -1.980 \\ -0.300 \\ -1.551
\end{tabular}
& \hspace{-0.5cm}
\begin{tabular}{c}
                     0.172 \\  0.172 \\  0.150 \\  0.157
\end{tabular}
& \hspace{-0.5cm}
\begin{tabular}{c}
                     1.141 \\  1.597 \\ -0.044 \\  0.952
\end{tabular}
& \hspace{-0.5cm}
\begin{tabular}{c}
                     0.201 \\  0.201 \\  0.351 \\  0.201
\end{tabular}
& \hspace{-0.5cm}
\begin{tabular}{c}
                     0.000 \\  0.000 \\ -0.245 \\  0.000
\end{tabular}
& \hspace{-0.5cm}
\begin{tabular}{c}
                     0.401 \\  0.401 \\  0.201 \\  0.401
\end{tabular}
\\ \hline
$(\gamma_{\mu} \otimes \gamma_{\mu})$ &
\begin{tabular}{c}
$I(c\bar c)$ \\ $II(c\bar c)$ \\  $U(c\bar c)$ \\ $R(c\bar c)$
\end{tabular}
& \hspace{-0.5cm}
\begin{tabular}{c}
                    -1.682 \\ -1.682 \\  1.000 \\ -1.312
\end{tabular}
& \hspace{-0.5cm}
\begin{tabular}{c}
                     0.000 \\  0.000 \\ -0.150 \\  0.000
\end{tabular}
& \hspace{-0.5cm}
\begin{tabular}{c}
                     1.225 \\  1.587 \\ -1.600 \\  0.977
\end{tabular}
& \hspace{-0.5cm}
\begin{tabular}{c}
                     0.201 \\  0.201 \\  0.201 \\  0.201
\end{tabular}
& \hspace{-0.5cm}
\begin{tabular}{c}
                     0.506 \\  0.445 \\  0.544 \\  0.492
\end{tabular}
& \hspace{-0.5cm}
\begin{tabular}{c}
                     0.001 \\  0.001 \\  0.001 \\  0.001
\end{tabular}
\\ \hline
\end{tabular}
\label{cs-2}
\ee
As concerns the solution  $U(c\bar c)$, we have included into
calculations the scalar and vector confinement forces, with
$b_S=-b_V=0.150$ GeV$^2$. The vector confinement potential is needed
for the description of the light quark  states ($q\bar q$) with large
masses. In this way, solution $U(c\bar c)$ gives us the description of
data with a universal confinement potential for all flavours.

The $\alpha_s$ coupling, being determined by the one-gluon exchange
forces, is of the same order in all solutions: $\alpha_s=3/4\cdot
d_V\simeq 0.38$.

The mass of the constituent $c$-quark is taken to be $m_c=1.25$
GeV. This mass value is consistent with the magnitude
provided by the heavy-quark effective
theory \cite{Isgur,Monohar}: $1.0\le m_c\le 1.4$ GeV; a slightly larger
interval for $m_c$ is given by lattice calculations,
$0.93\le m_c\le 1.59$ GeV, see
\cite{Monohar} and references therein. The compilation \cite{PDG} gives
us $1.15\le m_c\le 1.35$ GeV.

\subsection{Masses of $c\bar c$ states}

The fitting procedure results
in the following masses (in GeV units) for $1^{--}$ states:
\be
\begin{tabular}{lllllll}
State      & Data  & $I(c\bar c)$& $II(c\bar c)$& $U(c\bar c)$& $R(c\bar c)$ & $R^2_I$ \\
$J/\psi$   & {\bf3.097} & 3.100 $(S)$ & 3.132 $(S)$ & 3.115 $(S)$ & 3.109 $(S)$ &  2.060 \\
$\psi(2S)$ & {\bf3.686} & 3.676 $(S)$ & 3.662 $(S)$ & 3.635 $(S)$ & 3.681 $(S)$ &  6.897 \\
$\psi(1D)$ & {\bf3.770} & 3.794 $(D)$ & 3.770 $(D)$ & 3.747 $(D)$ & 3.788 $(D)$ &  2.060 \\
$\psi(3S)$ & {\bf4.040} & 4.079 $(S)$ & 4.049 $(S)$ & 4.009 $(S)$ & 4.107 $(S)$ & 12.636 \\
$\psi(2D)$ & {\bf4.160} & 4.156 $(D)$ & 4.121 $(D)$ & 4.087 $(D)$ & 4.168 $(D)$ &  6.897 \\
$\psi(4S)$ & {\bf4.415} & 4.434 $(S)$ & 4.386 $(S)$ & 4.290 $(S)$ & 4.498 $(S)$ & 17.227 \\
$\psi(3D)$ & ---        & 4.482 $(D)$ & 4.432 $(D)$ & 4.390 $(D)$ & 4.581 $(D)$ & 12.636 \\
$\psi(5S)$ & ---        & 4.781 $(S)$ & 4.737 $(S)$ & 4.566 $(S)$ & 4.812 $(S)$ & 32.968 \\
$\psi(4D)$ & ---        & 4.889 $(D)$ & 4.801 $(D)$ & 4.711 $(D)$ & 5.181 $(D)$ & 17.227 \\
$\psi(6S)$ & ---        & 5.135 $(S)$ & 5.057 $(S)$ & 4.993 $(S)$ & 5.429 $(S)$ & 23.372 \\
$\psi(5D)$ & ---        & 5.451 $(D)$ & 5.325 $(D)$ & 5.136 $(D)$ & 5.985 $(D)$ & 32.968 \\
$\psi(6D)$ & ---        & 6.030 $(D)$ & 5.869 $(D)$ & 5.819 $(D)$ & 6.943 $(D)$ & 23.372 \ ,
\end{tabular}
\label{cs-3}
\ee
Bold numbers stand for the masses included in the fit as an input.
The states $1^{--}$ are the mixture of $S$ and $D$ waves (in
parantheses the dominant wave is shown, see
indices $(nS)$ and $(nD)$). In Appendix, we present the wave functions
for the $1^{--}$ states and the values $W_{00}$, $W_{02}$, $W_{22}$,
which characterize a percentage of the $L=0$ and $L=2$ components (see
\cite{bb} for the details). The last column gives us the mean square
radii for the states in  solution $I(c\bar c )$: $R^2_I$ GeV$^{-2 }$.

For the other considered states, the fit resulted in the following
masses and  $R^2_I$ (all values in GeV units):

\noindent for $0^{-+}$ states $(L=0)$:
\be
\begin{tabular}{llllllll}
State        & Data & $I(c\bar c)$ & $II(c\bar c)$ & $U(c\bar c)$ & $R(c\bar c)$ & $R^2_I$ \\
$\eta_c(1S)$ & {\bf 2.979} & 2.979 & 2.985 & 3.016 & 2.970 &  1.682 \\
$\eta_c(2S)$ & {\bf 3.594} & 3.606 & 3.581 & 3.574 & 3.616 &  6.207 \\
$\eta_c(3S)$ & ---         & 4.030 & 3.994 & 3.958 & 4.067 & 11.813 \\
$\eta_c(4S)$ & ---         & 4.386 & 4.338 & 4.265 & 4.461 & 16.604 \\
$\eta_c(5S)$ & ---         & 4.763 & 4.715 & 4.555 & 4.813 & 30.919 \\
$\eta_c(6S)$ & ---         & 5.136 & 5.054 & 4.881 & 5.475 & 22.831 \ ,
\end{tabular}
\label{cs-4}
\ee
\noindent for $0^{++}$ states $(L=1)$:
\be
\begin{tabular}{lllllll}
State & Data & $I(c\bar c)$ & $II(c\bar c)$ & $U(c\bar c)$ & $R(c\bar c)$ & $R^2_I$ \\
$\chi_{c0}(1P)$ & {\bf 3.415} & 3.412 & 3.372 & 3.473 & 3.407 &  3.401 \\
$\chi_{c0}(2P)$ & ---         & 3.867 & 3.822 & 3.850 & 3.876 &  8.777 \\
$\chi_{c0}(3P)$ & ---         & 4.228 & 4.180 & 4.173 & 4.265 & 15.115 \\
$\chi_{c0}(4P)$ & ---         & 4.538 & 4.489 & 4.493 & 4.603 & 22.156 \\
$\chi_{c0}(5P)$ & ---         & 4.943 & 4.862 & 4.795 & 5.038 & 18.133 \\
$\chi_{c0}(6P)$ & ---         & 5.436 & 5.318 & 5.067 & 5.804 & 13.806 \ ,
\end{tabular}
\label{cs-5}
\ee
\noindent for $1^{++}$ states $(L=1)$:
\be
\begin{tabular}{lllllll}
State & Data  & $I(c\bar c)$ & $II(c\bar c)$ & $U(c\bar c)$ & $R(c\bar c)$ & $R^2_I$ \\
$\chi_{c1}(1P)$ & {\bf 3.510} & 3.500 & 3.491 & 3.503 & 3.497 &  4.234 \\
$\chi_{c1}(2P)$ & 3.872       & 3.933 & 3.904 & 3.880 & 3.944 &  9.861 \\
$\chi_{c1}(3P)$ & ---         & 4.278 & 4.242 & 3.989 & 4.328 & 17.628 \\
$\chi_{c1}(4P)$ & ---         & 4.563 & 4.524 & 4.228 & 4.809 & 24.460 \\
$\chi_{c1}(5P)$ & ---         & 4.916 & 4.841 & 4.575 & 5.635 & 18.407 \\
$\chi_{c1}(6P)$ & ---         & 5.460 & 5.345 & 4.819 & 6.445 & 13.345 \ ,
\end{tabular}
\label{cs-6}
\ee
\noindent for $1^{+-}$ states $(L=1)$:
\be
\begin{tabular}{lllllll}
State & Data & $I(c\bar c)$ & $II(c\bar c)$ & $U(c\bar c)$ & $R(c\bar c)$ & $R^2_I$ \\
$h_c(1P)$ & {\bf 3.526} & 3.511 & 3.513 & 3.522 & 3.515 &  4.447 \\
$h_c(2P)$ & ---         & 3.949 & 3.918 & 4.013 & 3.942 & 10.199 \\
$h_c(3P)$ & ---         & 4.323 & 4.260 & 4.385 & 4.300 & 14.886 \\
$h_c(4P)$ & ---         & 4.678 & 4.599 & 4.696 & 4.659 & 19.976 \\
$h_c(5P)$ & ---         & 5.158 & 4.916 & 5.078 & 4.997 & 24.106 \\
$h_c(6P)$ & ---         & 5.968 & 5.353 & 5.531 & 5.472 & 15.336 \ ,
\end{tabular}
\label{cs-7}
\ee
\noindent and for $2^{++}$ states $(L=1,3)$:
\be
\begin{tabular}{lllllllll}
State       &          Data & $I(c\bar c)$ & $II(c\bar c)$& $U(c\bar c)$& $R(c\bar c)$ & $R^2_I$ \\
$\chi_{c2}(1P)$ & {\bf 3.556}   & 3.552 $(P)$ & 3.570 $(P)$ & 3.508 $(P)$ & 3.556 $(P)$ &  5.008 \\
$\chi_{c2}(2P)$ & 3.941         & 3.986 $(P)$ & 3.957 $(P)$ & 3.898 $(P)$ & 3.969 $(P)$ & 11.085 \\
$\chi_{c2}(1F)$ & ---           & 4.050 $(F)$ & 4.035 $(F)$ & 3.946 $(F)$ & 4.054 $(F)$ &  5.008 \\
$\chi_{c2}(3P)$ & ---           & 4.350 $(P)$ & 4.288 $(P)$ & 4.222 $(P)$ & 4.320 $(P)$ & 14.928 \\
$\chi_{c2}(2F)$ & ---           & 4.476 $(F)$ & 4.333 $(F)$ & 4.260 $(F)$ & 4.362 $(F)$ & 11.085 \\
$\chi_{c2}(4P)$ & ---           & 4.786 $(P)$ & 4.402 $(P)$ & 4.546 $(P)$ & 4.411 $(P)$ & 41.793 \\
$\chi_{c2}(3F)$ & ---           & 5.073 $(F)$ & 4.648 $(F)$ & 4.558 $(F)$ & 4.705 $(F)$ & 14.928 \\
$\chi_{c2}(5P)$ & ---           & 5.505 $(P)$ & 4.687 $(P)$ & 4.803 $(P)$ & 4.762 $(P)$ & 12.018 \\
$\chi_{c2}(4F)$ & ---           & 5.833 $(F)$ & 5.055 $(F)$ & 4.937 $(F)$ & 5.153 $(F)$ & 41.793 \\
$\chi_{c2}(6P)$ & ---           & 6.318 $(P)$ & 5.223 $(P)$ & 5.079 $(P)$ & 5.334 $(P)$ & 10.590 \\
$\chi_{c2}(5F)$ & ---           & 6.765 $(F)$ & 5.596 $(F)$ & 5.429 $(F)$ & 5.727 $(F)$ & 12.018 \\
$\chi_{c2}(6F)$ & ---           & 7.710 $(F)$ & 6.154 $(F)$ & 6.065
$(F)$ & 6.325 $(F)$ & 10.590 \ ,
 \end{tabular}
\label{cs-8a}
\ee
\noindent for
$2^{-+}$ states $(L=2)$:
\be
\begin{tabular}{lllllll}
State & Data &
$I(c\bar c)$ & $II(c\bar c)$ & $U(c\bar c)$ & $R(c\bar c)$ & $R^2_I$ \\
$\eta_{c2}(1D)$ & ---         & 3.829 & 3.823 & 3.742 & 3.818 &  7.721 \\
$\eta_{c2}(2D)$ & ---         & 4.185 & 4.164 & 4.087 & 4.185 & 14.387 \\
$\eta_{c2}(3D)$ & ---         & 4.471 & 4.449 & 4.397 & 4.591 & 22.729 \\
$\eta_{c2}(4D)$ & ---         & 4.788 & 4.745 & 4.713 & 5.151 & 18.708 \\
$\eta_{c2}(5D)$ & ---         & 5.242 & 5.157 & 5.084 & 5.996 & 14.024 \\
$\eta_{c2}(6D)$ & ---         & 5.813 & 5.693 & 5.546 & 6.832 & 12.227
\ .
\end{tabular}
\label{cs-8b}
\ee

\begin{figure}
\centerline{\epsfig{file=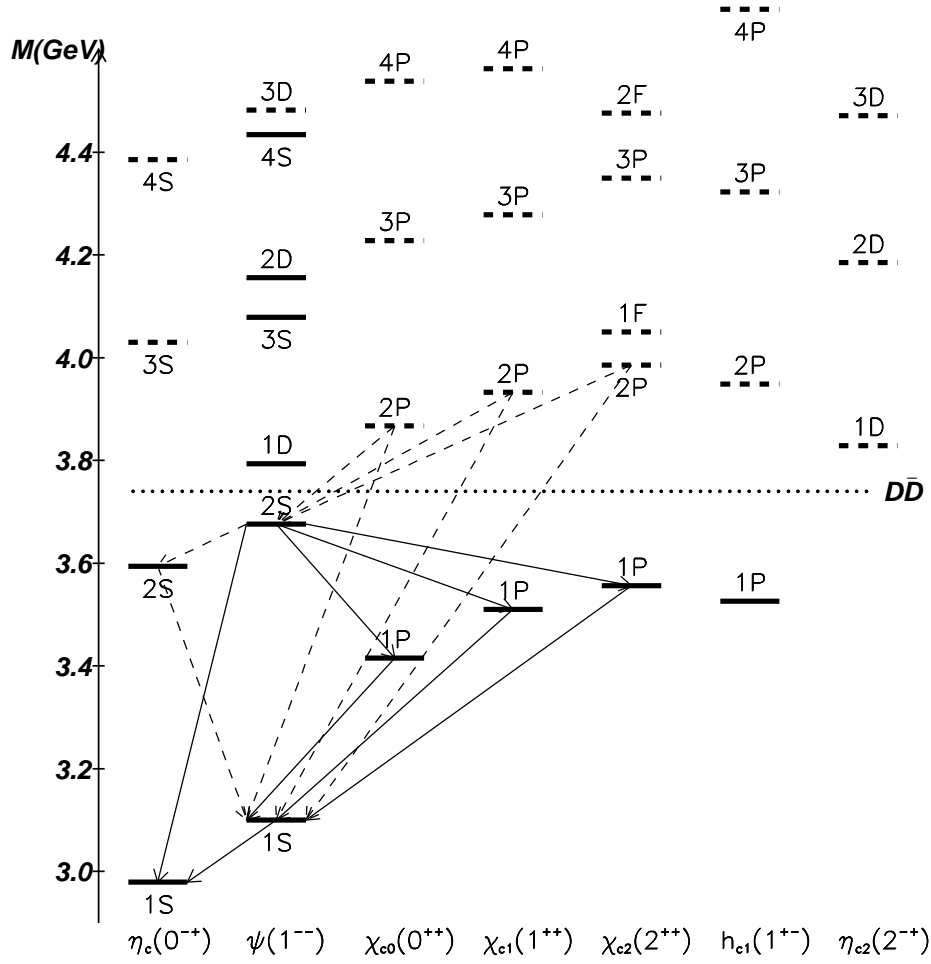,width=13cm}}
\vspace{1.0cm}
\caption{ The $c\bar c$ levels in solution $I(c\bar c)$. Solid lines
stand
for observed states, dashed lines for  the predicted ones. Thin solid
lines show the transitions  included into fitting procedure,
thin dashed line demonstrates the transitions, whose widths  are
predicted.}
 \end{figure}

In Fig. 1, the levels for solution
$I(c\bar c)$ are shown for the mass region  $M < 4.5$ GeV. The wave
functions for this solution are given in Appendix.

The obtained variants of solutions allow us to consider the status
of $X(3872)$ and $Y(3941)$.

The solution $II(c\bar c)$ provides for $\chi_{c1}(2P)$ the mass
$3904$ MeV that is close to the value found in
\cite{Choi,Acosta,Abazov,Aubert} for the pick denoted as $X(3872)$.
In agreement with $II(c\bar c)$-solution, the analysis
\cite{Bugg} favours the  quantum numbers $J^{PC}=1^{++}$ for this
state.

Still, one cannot exclude that  $X(3872)$ is the
$2^{-+}$-state (see \cite{Barnes} and references therein).  In
all solutions, the mass of $\eta_{c2}(1D)$ is lying in the interval
$3815-3830$ MeV that demonstrates the plausibility of this variant as
well.

The signal $Y(3941)$ does not contradict the hypothesis about
its $2^{++}$ nature, because  the mass of
the $\chi_{2c}(2P)$ is close to $Y(3941)$ in all solutions.

\subsection{Radiative transitions $(c\bar c)_{in} \to\gamma +(c\bar
c)_{out}$}

In Fig. 1, we show radiative decays which have been accounted for
in the fitting procedure (corresonding formulae are presented in
\cite{bb,raddecay}). For the levels below $D\bar D$ threshold,
experimental data
\cite{PDG,c1-1,c1-2,Hernandez}
and the magnitudes of widths
obtained in different versions of the fit are as follows (in keV):
 \be
\begin{tabular}{llllll}
Process & Data & $I(c\bar c)$ & $II(c\bar c)$ & $U(c\bar c)$ & $R(c\bar c)$ \\
$J/\psi\to \gamma\eta_{c0} (1S)$  & 1.1$\pm$0.3     &   2.877 &   4.853 &   1.4   &   4.216 \\
$\chi_{c0}(1P)\to\gamma J/\psi$   & 165$\pm$50      & 180.845 & 164.884 & 273.8   & 180.578 \\
$\chi_{c1}(1P)\to\gamma J/\psi$   & 295$\pm$90      & 291.536 & 297.362 & 391.8   & 289.569 \\
$\chi_{c2}(1P)\to\gamma J/\psi$   & 390$\pm$120     & 226.700 & 239.835 & 312.3   & 225.265 \\
$\eta_{c0}(2S)\to\gamma J/\psi$   & ---             &  23.297 &  23.326 &  40.263 &  25.852 \\
$\psi (2S) \to \gamma\eta_{c0}(1S)$   & 0.8$\pm$0.2 &   1.661 &   2.529 &   0.37  &   2.309 \\
$\psi (2S) \to \gamma\chi_{c0}(1P)$   & 26$\pm$4    &  22.854 &  32.408 &  12.2   &  22.169 \\
$\psi (2S) \to \gamma\chi_{c1}(1P)$   & 25$\pm$4    &  45.291 &  51.853 &  31.1   &  44.343 \\
$\psi (2S) \to \gamma\chi_{c2}(1P)$   & 20$\pm$4    &  26.096 &  20.929 &  40.2   &  26.620 \\
$\psi (2S) \to \gamma\eta_{c0}(2S)$   & ---         &   0.541 &   0.770 &   1.003 &   0.447 \\
\end{tabular}
\label{cs-9}
\ee
Note that the 20\%
accuracy is allowed for the transitions
$\psi (2S)\to\gamma \chi_{cJ}(1P)$ and 30\%
one for $ \chi_{cJ}(1P)\to\gamma\psi (1S)$ (we use a bit larger
errors than those
obtained in \cite{Hernandez}).
We also  predict the widths of  the decays
$\eta_{c0}\to\gamma J/\psi$ and $\psi(2S)\to\gamma\eta_{c0}(2S)$.

The calculated values in (\ref{cs-9}) agree rather reasonably
with the  data.

The predictions of widths  of the levels above the $D\bar D$
threshold (see Fig. 1) are as follows (widths are in keV):
\be
\begin{tabular}{cccccc}
Process & Data        & $I(c\bar c)$ & $II(c\bar c)$ & $U(c\bar c)$ & $R(c\bar c)$ \\
$\chi_{c0}(2P)\to\gamma J/\psi$   & --- &   9.181 &   5.841 &   0.468 &  12.612 \\
$\chi_{c1}(2P)\to\gamma J/\psi$   & --- &  48.130 &  50.831 &  28.797 &  59.013 \\
$\chi_{c2}(2P)\to\gamma J/\psi$   & --- &  56.911 &  72.080 &  31.331 &  67.713 \\
$\chi_{c0}(2P)\to\gamma \psi(2S)$ & --- &  81.851 &  73.158 &  92.450 &  73.925 \\
$\chi_{c1}(2P)\to\gamma \psi(2S)$ & --- & 212.937 & 221.604 & 290.379 & 192.502 \\
$\chi_{c2}(2P)\to\gamma \psi(2S)$ & --- & 154.796 & 176.599 & 197.162 & 144.714 \\
\label{cs-9a}
\end{tabular}
\ee

\subsection{The $c\bar c $ component of the photon wave
function and two-photon radiative decays}

In the fitting procedure, we approximate the vertex of the transition
$\gamma\to c\bar c$ by the following formula:
\bea
&&G_{\gamma\to c\bar c(S)}(s)=
\sum\limits_{n=1}^6 C_{nS} G_{V(nS)}(s) +
\frac1{1+\exp(-\beta_\gamma (s-s_0))}\ ,
\\ \nn
&&G_{\gamma\to c\bar c(D)}(s)=
\sum\limits_{n=1}^2 C_{nD} G_{V(nD)}(s)\ ,
\label{cs-10}
\eea
where $G_{V(nS)}(s)$
is the vertex for the transition $\psi(nS)\to c\bar c$ and
$G_{V(nD)}(s)$
is the vertex for the transition $\psi(nD)\to c\bar c$, see
\cite{bb} for the details. The parameters
$C_{nS}$, $C_{nD}$, $\beta_\gamma$, $s_0$
for solutions  $I(c\bar c)$, $II(c\bar c)$, $U(c\bar c)$  and
$R(c\bar c)$ have been found as follows (in GeV):

\begin{equation}
\begin{tabular}{lllll}
$I(c\bar c)$       & $II(c\bar c)$       & $U(c\bar c)$         & $R(c\bar c)$       \\
$C_{1S}$ = -10.100 & $C_{1S}$ = -9.073   & $C_{1S}$ = -3.852    & $C_{1S}$ = -14.990 \\
$C_{2S}$ =   2.121 & $C_{2S}$ = -0.252   & $C_{2S}$ =  0.476    & $C_{2S}$ =   3.416 \\
$C_{3S}$ =  -0.373 & $C_{3S}$ =  0.038   & $C_{3S}$ =  0.325    & $C_{3S}$ =  -0.862 \\
$C_{4S}$ =  -1.818 & $C_{4S}$ =  0.217   & $C_{4S}$ =  0.667    & $C_{4S}$ =  -1.834 \\
$C_{5S}$ =   9.998 & $C_{5S}$ = -1.158   & $C_{5S}$ = -2.571    & $C_{5S}$ =  11.704 \\
$C_{6S}$ =   3.369 & $C_{6S}$ = -0.529   & $C_{6S}$ = -0.707    & $C_{6S}$ =   0.305 \\
$C_{1D}$ =  -0.031 & $C_{1D}$ = -0.005   & $C_{1D}$ =  0.080    & $C_{1D}$ =  -0.186 \\
$C_{2D}$ =  -0.110 & $C_{2D}$ =  0.020   & $C_{2D}$ = -0.082    & $C_{2D}$ =   0.114 \\
$b_\gamma$ = 2.85  & $b_\gamma$ = 2.85   & $b_\gamma$ = 2.85    & $b_\gamma$ = 2.85  \\
$s_0$ =    18.79   & $s_0$ =    18.79    & $s_0$ =    18.79     & $s_0$ =    18.79
\end{tabular}
\label{cs-11}
\end{equation}
The corresponding vertices
$G_{\gamma\to c\bar c}(s)$ are shown in Fig. 2.

\begin{figure}
\centerline{\epsfig{file=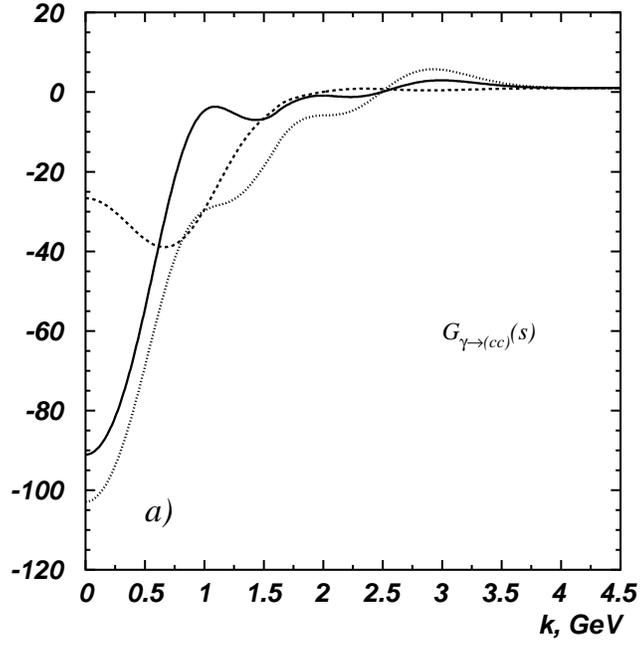,width=9cm}}
\centerline{\epsfig{file=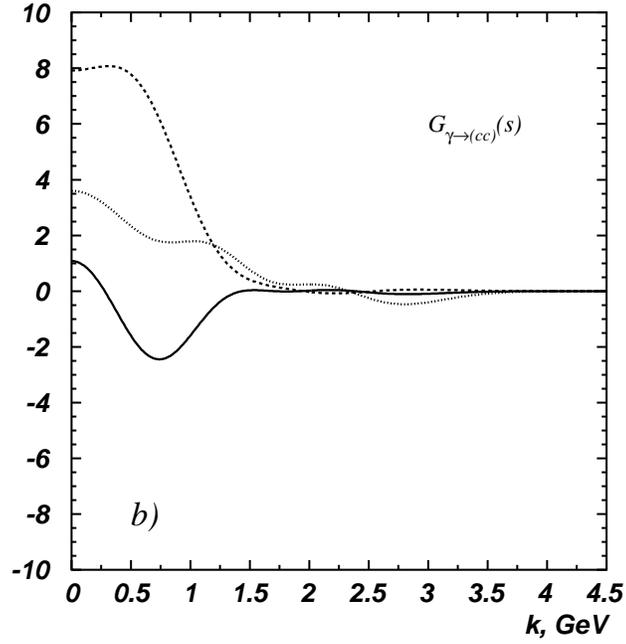,width=9cm}}
\caption{
a) Vertices $G^{(S)}_{\gamma\to c\bar c}$ (a) and
$G^{(D)}_{\gamma\to c\bar c}$ (b)
for solutions $I(c\bar c)$ (solid lines), $U(c\bar c)$ (dashed lines)
and $R(c\bar c)$ (dotted lines).}
\end{figure}

Experimental values of partial widths \cite{PDG,L3,OPAL,CLEO,E760}
 together with the widths
obtained in the fitting procedure
are shown below (in keV):
\be
\begin{tabular}{cccccc}
Process & Data        & $I(c\bar c)$ & $II(c\bar c)$ & $U(c\bar c)$ & $R(c\bar c)$ \\
$J/\psi(1S)\to e^+e^-$ & 5.40 $\pm$ 0.22 & 5.392 & 5.360 & 5.403 & 6.088 \\
  $\psi(2S)\to e^+e^-$ & 2.14 $\pm$ 0.21 & 2.146 & 2.137 & 2.142 & 2.843 \\
  $\psi(1D)\to e^+e^-$ & 0.24 $\pm$ 0.05 & 0.239 & 0.730 & 0.240 & 0.301 \\
  $\psi(3S)\to e^+e^-$ & 0.75 $\pm$ 0.15 & 0.754 & 0.472 & 0.749 & 1.210 \\
  $\psi(2D)\to e^+e^-$ & 0.47 $\pm$ 0.10 & 0.472 & 0.236 & 0.469 & 0.113 \\
  $\psi(4S)\to e^+e^-$ & 0.77 $\pm$ 0.23 & 0.771 & 0.763 & 0.770 & 0.915 \\
\label{cs-12}
\end{tabular}
\ee

With the determined vertices for
$G_{\gamma\to c\bar c}(s)$, we can obtain the widths of the two-photon
decays (see \cite{bb} for more detail). The comparison of
experimentally measured widths with those obtain in our calculations
is given below:
\be
\begin{tabular}{cccccc}
Process & Data        & $I(c\bar c)$ & $II(c\bar c)$ & $U(c\bar c)$ & $R(c\bar c)$ \\
$\eta_{c0} (1S)\to\gamma\gamma$ & 7.0$\pm$0.9                  & 6.971 & 6.808 & 7.002 & 7.088 \\
$\chi_{c0} (1P)\to\gamma\gamma$ & 2.6$\pm$0.5                  & 2.426 & 2.620 & 2.578 & 2.212 \\
$\chi_{c2} (1P)\to\gamma\gamma$ & 1.02$\pm$0.40$\pm$0.17(L3  ) & 0.558 & 0.528 & 0.068 & 5.706 \\
                                & 1.76$\pm$0.47$\pm$0.40(OPAL) & &  &   \\
                                & 1.08$\pm$0.30$\pm$0.26(CLEO) & &  &   \\
                                & 0.33$\pm$0.08$\pm$0.06(E760) & &  &   \\
\label{cs-13}
\end{tabular}
\ee
Let us emphasize that the data do not tell us definitely about the
width $\chi_{c2}(3556)\to \gamma\gamma$. In the
reaction $p\bar p\to \gamma\gamma$, the value $\Gamma(\chi_2(3556)\to
\gamma\gamma)=0.32\pm 0.080\pm 0.055$ keV was obtained in \cite{E760},
while in direct measurements  such as $e^+e^-$ annihilation
the width is much larger:
$1.02\pm 0.40\pm 0.17\;{\rm keV}$ \cite{L3}\ ,
$1.76\pm 0.47\pm 0.40\;{\rm keV}$ \cite{OPAL}\  ,
$1.08\pm 0.30\pm 0.26 \;{\rm keV}$ \cite{CLEO}\  .
The compilation \cite{PDG} provides us with the value
close to  that of \cite{E760}. The value found in our fit agrees with
data reported by
\cite{L3,OPAL,CLEO}
and contradicts the magnitude from
\cite{E760}.

Our predictions of widths $c\bar c\to\gamma\gamma$ for the levels
below $4$ GeV are as follows:
\be
\begin{tabular}{cccccc}
Process & Data        & $I(c\bar c)$ & $II(c\bar c)$ & $U(c\bar c)$ & $R(c\bar c)$ \\
$\eta_{c0} (2S)\to\gamma\gamma$ & --- & 12.234 & 11.947 & 12.289 & 12.439 \\
$\chi_{c0} (2P)\to\gamma\gamma$ & --- &  2.142 &  2.313 &  2.276 &  1.953 \\
$\chi_{c2} (2P)\to\gamma\gamma$ & --- &  0.500 &  0.473 &  0.061 &  5.111 \\
\label{cs-14}
\end{tabular}
\ee
In tables 1,2,3, we compare our results with those obtained by  other
authors.

In \cite{Linde}, the calculated widths
depend on a chosen gauge for the gluon exchange interaction
--- we demonstrate the results obtained
 for  both Feynman (F) and Coulomb (C) gauges.

In \cite{Resag}, the $c\bar c$ system was studied in terms of scalar (S)
and vector (V) confinement forces ---
both variants are presented in tables 1,2,3. The results
obtained in the nonrelativistic approach to the $c\bar c$
system  \cite{NR} are also shown in tables 1,2,3.

In both  relativistic \cite{Linde,Resag} and nonrelativistic
\cite{NR} approaches, there is rather large discrepancy between the
data and calculated values of $\psi(nS)\to e^+e^-$ (in \cite{Linde} the
width of the transition $J/\psi\to e^+e^-$ was fixed with the use of a
subtraction parameter).
In our opinion, the reason is that in all
above-mentioned papers, soft interaction of quarks
was not accounted for --- we mean the processes shown in Fig. 3b,c of
Ref. \cite{bb}.
In fact, the necessity of taking into consideration the low-energy quark
interaction was understood decades ago
 but untill now this procedure has not become commonly
accepted even for light quarks; see, for example,
\cite{deWitt,Xiao}.

\begin{table}[h]
\caption{Comparison of  data for the transitions
$(c\bar c)_{in}\to\gamma + (c\bar c)_{out}$
with our results and calculations of other groups (the width is given
in keV).}
\begin{center} {\scriptsize
\begin{tabular}{|l|c|c|c|c|c|c|c|} \hline Decay & Data & $I(c\bar c)$ &
LS(F)\cite{Linde} & LS(C)\cite{Linde}
&RM(S)\cite{Resag}&RM(V)\cite{Resag}&NR\cite{NR}\\
\hline
$J/\psi (1S)\to\eta_{c0}(1S)\gamma$ & 1.1$\pm$0.3 & 3.4 & 1.7--1.3 & 1.7--1.4 & 3.35 & 2.66 & 1.21 \\
\hline
$\psi(2S)\to\chi_{c0}(1P)\gamma$ & 26$\pm$4 & 23 & 31--47 & 26--31
 & 31 & 32 & 19.4 \\
$\psi(2S)\to\chi_{c1}(1P)\gamma$ & 25$\pm$4 & 45 & 58--49 & 63--50
 & 36 & 48 & 34.8 \\
$\psi(2S)\to\chi_{c2}(1P)\gamma$ & 20$\pm$4 & 26 & 48--47 & 51--49
& 60 & 35 & 29.3 \\
\hline
$\psi(2S)\to\eta_{c0}(1S)\gamma$ & 0.8$\pm$0.2 & 2.0 & 11--10 & 10--13
& 6 & 1.3 & 4.47\\
\hline
$\chi_{c0}(1P)\to J/\psi(1S)\gamma$ & 165$\pm$50 & 181 & 130--96
& 143--110 & 140 & 119 & 147 \\
$\chi_{c1}(1P)\to J/\psi(1S)\gamma$ & 295$\pm$90 & 292 & 390--399
& 426--434 & 250 & 230 & 287 \\
$\chi_{c2}(1P)\to J/\psi(1S)\gamma$ & 390$\pm$120& 227 & 218--195
 & 240--218 & 270 & 347 & 393 \\
\hline
\end{tabular}
}
\end{center}
\end{table}

\begin{table}[h]
\caption{Decay widths
$\psi\to e^+e^-$, our results and those of other groups v.s. data (in
keV).}
\begin{center} {\scriptsize
\begin{tabular}{|l|c|c|c|c|c|c|c|} \hline Decay & Data & $I(c\bar c)$ &
LS(F)\cite{Linde} & LS(C)\cite{Linde}
&RM(S)\cite{Resag}&RM(V)\cite{Resag}&NR\cite{NR}\\
\hline
$J/\psi(1S)\to e^+e^-$ & 5.40 $\pm$ 0.22 & 5.39 & 5.26     & 5.26     & 8.05 & 9.21 & 12.2 \\
$  \psi(2S)\to e^+e^-$ & 2.14 $\pm$ 0.21 & 2.15 & 2.8--2.5 & 2.9--2.7 & 4.30 & 5.87 & 4.63 \\
$  \psi(1D)\to e^+e^-$ & 0.24 $\pm$ 0.05 & 0.24 & 2.0--1.6 & 2.1--1.8 & 3.05 & 4.81 & 3.20 \\
$  \psi(3S)\to e^+e^-$ & 0.75 $\pm$ 0.15 & 0.75 & 1.4--1.0 & 1.6--1.3 & 2.16 & 3.95 & 2.41 \\
$  \psi(2D)\to e^+e^-$ & 0.47 $\pm$ 0.10 & 0.47 & ---      & ---      & ---  & ---  & ---  \\
$  \psi(4S)\to e^+e^-$ & 0.77 $\pm$ 0.23 & 0.77 & ---      & ---      & ---  & ---  & ---  \\
\hline
\end{tabular}
}
\end{center}
\end{table}

\begin{table}[h]
\caption{Decay widhs $(c\bar c)_{in}\to\gamma + (c\bar c)_{out}$
in the region below $4$ GeV (in keV)}
\begin{center}
 {\scriptsize
\begin{tabular}{|l|c|c|c|c|c|c|c|c|c|}
\hline
Decay & Data & $I(c\bar c)$ & LS \cite{Linde} &
\cite{G} & \cite{M} & \cite{Gupta} & \cite{Sch} & \cite{Huang} &
\cite{Barnes2}\\
 \hline $\eta_{c }(1S)\to\gamma\gamma$ & 7.0$\pm$0.9 &
6.9 & 6.2--6.3 (F,C) &
 5.5 & 3.5  & 10.9 & 7.8 & 5.5 & 4.8 \\
$\eta_{c }(1S)\to\gamma\gamma$ & ---         & 12.2 & --             &
 1.8 & 1.38 & --   & 3.5 & 2.1 & 3.7 \\
\hline
$\chi_{c0}(1P)\to\gamma\gamma$ & 2.6$\pm$0.5 &  2.4 & 1.5--1.8 (F,C) &
 2.9 & 1.39 & 6.4 & 2.5 & 5.32 & --  \\
$\chi_{c0}(1P)\to\gamma\gamma$ & ---         &  2.1 & ---            &
 1.9 & 1.11 & --  & --  & --   & --  \\
\hline
$\chi_{c2}(1P)\to\gamma\gamma$ & 1.02$\pm$0.40$\pm$0.17\cite{L3}   &
0.61 & 0.3--0.4 (F,C)& 0.50 & 0.44 & 0.57 &0.28 & 0.44 & -- \\
                               & 1.76$\pm$0.47$\pm$0.40\cite{OPAL} &
 &               &      & & & & & \\
                               & 1.08$\pm$0.30$\pm$0.26\cite{CLEO} &
 &               &      & & & & & \\
                               & 0.33$\pm$0.08$\pm$0.06\cite{E760} &
 &               &      & & & & & \\
$\chi_{c2}(1P)\to\gamma\gamma$ & ---                               & 0.5
 & ---           & 0.52 & 0.48 & --   & --  & --   & -- \\
\hline
\end{tabular}
}
\end{center}
\end{table}

\section{Conclusion}

We have performed a successful description of both $c\bar c$ levels and
their radial excitation transitions for several interaction variants as
follows:
for scalar confinement potential, scalar and vector confinement
potential in case of instantaneous and retardation forces.
Such a diversity in the choice of  forces reflects the lack of
available experimental data to guarantee unambiguous determination of
the interaction. First of all, one misses the data for the radiative
$c\bar c$ decays. This is why we pay special attention to the
predictions given by different versions of our calculations.

Rather good description of the $c\bar c$ levels being obtained, the wave
functions related to different solutions may noticeably differ from each
other, and sometimes they demonstrate rather unusual behaviour. Because
of that, we pay a considerable attention to the presentation of wave
functions of the $c\bar c$ systems. We think  that the future progress
in understanding the $c$-quark interactions in the soft region is
related precisely  to a better knowledge of wave functions, that
should be based on further study of radiative transitions.

\section*{Acknowledgments}

We thank A.V. Anisovich, Y.I. Azimov, G.S. Danilov, I.T. Dyatlov,
L.N. Lipatov, V.Y. Petrov, H.R. Petry and M.G. Ryskin
for useful discussions.

This work was supported by the Russian Foundation for Basic Research,
project no. 04-02-17091.

\section{Appendix. Wave function of the $c\bar c$ sector}{**}

Tables 4--10 give us the $c_i(S,L,J;n)$ coefficients,
which determine the wave functions $\psi^{(S,L,J)}$ for
solution $I(c\bar c)$ according to the following formula:
\bea
\nonumber
\label{a-1}
\psi^{(S,L,J)}_{(n)}(k^2)=e^{-\beta k^2}\sum\limits_{i=1}^9 c_i(S,L,J;n)
k^{i-1}\, ,
\eea
where $ k^2\equiv {\bf k}^2 $
(recall that $s=4m^2+4{\bf k}^2$). The
fitting parameter $\beta$ is of the order of $0.5-1.5 $ GeV$^{-2}$ and
it may be different in different flavor sectors; we put $\beta =1.2$
GeV$^{-2}$. In tables, we also
show $W_{LL'}$'s, which enter the normalization condition
$W_{LL}+W_{LL'}+W_{L'L'}=1$, being the convolution of
the wave functions
$$ W_{L,L'}=\psi^{(S,L,J)}\otimes\psi^{(S,L',J)}\ ,  $$
 see \cite{bb} for more detail.
 In Figs. 3--9, we
demonstrate these wave functions.

\begin{table}
\caption{Constants $c_i(S,L,J;n)$ from Eq. (\ref{a-1}) (in GeV units) for
the wave functions of $\Upsilon$-mesons in solution $I(c\bar c)$}

\begin{center} {\scriptsize
\begin{tabular}{|r|r|r|r|r|r|r|}
\hline
& \multicolumn{2}{|c|}{$\Psi(1S)$} & \multicolumn{2}{|c|}{$\Psi(2S)$} &
\multicolumn{2}{|c|}{$\Psi(3S)$} \\
\hline
 & \multicolumn{2}{|c|}{$W_{00}\qquad W_{02}\qquad W_{22}$} &
   \multicolumn{2}{|c|}{$W_{00}\qquad W_{02}\qquad W_{22}$} &
   \multicolumn{2}{|c|}{$W_{00}\qquad W_{02}\qquad W_{22}$} \\
 & \multicolumn{2}{|c|}{0.99987 \, -0.00066 \, 0.00079} &
   \multicolumn{2}{|c|}{0.99924 \, -0.00174 \, 0.00250} &
   \multicolumn{2}{|c|}{0.99807 \, -0.00288 \, 0.00481} \\
\hline
$i$ & $\psi^{(1,0,1)}$ & $\psi^{(1,2,1)}$ & $\psi^{(1,0,1)}$ & $\psi^{(1,2,1)}$  & $\psi^{(1,0,1)}$ & $\psi^{(1,2,1)}$  \\
\hline
 1 &         7.9009 &        -5.9190 &        11.6473 &         1.5241 &        17.3698 &       -30.6053 \\
 2 &        -6.0655 &        34.4862 &        47.6240 &       -14.8598 &        41.6854 &       199.5601 \\
 3 &        16.7028 &       -83.5239 &      -453.2935 &        41.5693 &      -860.2544 &      -499.8739 \\
 4 &      -105.4498 &       102.1413 &      1094.1453 &       -41.5148 &      2693.3655 &       625.7537 \\
 5 &       224.9337 &       -63.9523 &     -1334.8514 &         2.7143 &     -3793.6194 &      -415.9994 \\
 6 &      -231.2288 &        16.2702 &       943.6424 &        24.1754 &      2870.7394 &       134.3492 \\
 7 &       127.1267 &         1.9716 &      -393.8494 &       -18.1178 &     -1206.4102 &        -9.1646 \\
 8 &       -36.0487 &        -1.8633 &        90.4811 &         5.3507 &       264.5053 &        -5.3519 \\
 9 &         4.1617 &         0.2684 &        -8.8655 &        -0.5785 &       -23.4715 &         0.9886 \\
\hline
\hline
& \multicolumn{2}{|c|}{$\Psi(4S)$} & \multicolumn{2}{|c|}{$\Psi(5S)$} &
  \multicolumn{2}{|c|}{$\Psi(6S)$} \\
\hline
 & \multicolumn{2}{|c|}{$W_{00}\qquad W_{02}\qquad W_{22}$} &
   \multicolumn{2}{|c|}{$W_{00}\qquad W_{02}\qquad W_{22}$} &
   \multicolumn{2}{|c|}{$W_{00}\qquad W_{02}\qquad W_{22}$} \\
 & \multicolumn{2}{|c|}{0.99166 \, -0.00737 \, 0.01570} &
   \multicolumn{2}{|c|}{0.99889 \, -0.00450 \, 0.00561} &
   \multicolumn{2}{|c|}{0.95583 \, -0.02419 \, 0.06835} \\
\hline
$i$ & $\psi^{(1,0,1)}$ & $\psi^{(1,2,1)}$ & $\psi^{(1,0,1)}$ & $\psi^{(1,2,1)}$  & $\psi^{(1,0,1)}$ & $\psi^{(1,2,1)}$  \\
\hline
 1 &        36.5862 &        -1.1743 &       -95.0826 &        -9.0172 &        78.7073 &        62.0712 \\
 2 &      -212.7589 &        51.7345 &      1136.7086 &        81.4929 &     -1113.5476 &      -481.4354 \\
 3 &        -6.8022 &      -243.2839 &     -4872.0833 &      -284.5462 &      5651.5790 &      1419.5113 \\
 4 &      1855.8350 &       448.2876 &     10215.8015 &       523.5176 &    -13997.7499 &     -2062.1487 \\
 5 &     -4206.0402 &      -401.5179 &    -11814.1052 &      -570.3848 &     18993.0373 &      1573.0804 \\
 6 &      4182.8669 &       178.7153 &      7907.6179 &       381.1356 &    -14736.1180 &      -594.8321 \\
 7 &     -2151.4758 &       -31.8182 &     -3052.3529 &      -152.6765 &      6486.3763 &        71.5608 \\
 8 &       559.1998 &        -1.4693 &       631.1276 &        33.4538 &     -1501.0222 &        15.2345 \\
 9 &       -58.1665 &         0.8430 &       -54.2492 &        -3.0632 &       141.2673 &        -3.6824 \\
\hline
\hline
& \multicolumn{2}{|c|}{$\Psi(1D)$} & \multicolumn{2}{|c|}{$\Psi(2D)$} &
  \multicolumn{2}{|c|}{$\Psi(3D)$} \\
\hline
 & \multicolumn{2}{|c|}{$W_{00}\qquad W_{02}\qquad W_{22}$} &
   \multicolumn{2}{|c|}{$W_{00}\qquad W_{02}\qquad W_{22}$} &
   \multicolumn{2}{|c|}{$W_{00}\qquad W_{02}\qquad W_{22}$} \\
 & \multicolumn{2}{|c|}{0.00435 \, -0.00169 \, 0.99734} &
   \multicolumn{2}{|c|}{0.00736 \, -0.00316 \, 0.99581} &
   \multicolumn{2}{|c|}{0.00697 \, -0.00433 \, 0.99736} \\
\hline
$i$ & $\psi^{(1,2,1)}$ & $\psi^{(1,0,1)}$ & $\psi^{(1,2,1)}$ & $\psi^{(1,0,1)}$  & $\psi^{(1,2,1)}$ & $\psi^{(1,0,1)}$  \\
\hline
 1 &        84.4242 &        -0.1030 &       313.1175 &        -0.2904 &       467.5023 &        -1.8858 \\
 2 &      -205.1653 &         3.1892 &     -1319.9899 &        10.2284 &     -2389.2054 &        23.4791 \\
 3 &       181.7717 &       -19.8052 &      2113.8219 &       -70.5461 &      4394.0661 &      -121.9462 \\
 4 &       -77.1585 &        46.4276 &     -1592.4362 &       192.4214 &     -3418.8591 &       332.6430 \\
 5 &        49.8493 &       -58.9700 &       460.0515 &      -260.8978 &       524.5556 &      -507.1882 \\
 6 &       -57.6810 &        44.8041 &       111.8270 &       194.4329 &       907.2750 &       442.0815 \\
 7 &        35.7092 &       -20.4094 &      -120.5831 &       -80.7753 &      -629.5731 &      -218.2705 \\
 8 &       -10.1511 &         5.1363 &        32.7133 &        17.4329 &       163.7261 &        56.6805 \\
 9 &         1.0906 &        -0.5496 &        -3.1245 &        -1.5051 &       -15.6276 &        -6.0100 \\
\hline
\hline
& \multicolumn{2}{|c|}{$\Psi(4D)$} & \multicolumn{2}{|c|}{$\Psi(5D)$} &
  \multicolumn{2}{|c|}{$\Psi(6D)$} \\
\hline
 & \multicolumn{2}{|c|}{$W_{00}\qquad W_{02}\qquad W_{22}$} &
   \multicolumn{2}{|c|}{$W_{00}\qquad W_{02}\qquad W_{22}$} &
   \multicolumn{2}{|c|}{$W_{00}\qquad W_{02}\qquad W_{22}$} \\
 & \multicolumn{2}{|c|}{0.02465 \, -0.01134 \, 0.98669} &
   \multicolumn{2}{|c|}{0.01698 \, -0.01676 \, 0.99978} &
   \multicolumn{2}{|c|}{0.02212 \, -0.01885 \, 0.99673} \\
\hline
$i$ & $\psi^{(1,2,1)}$ & $\psi^{(1,0,1)}$ & $\psi^{(1,2,1)}$ & $\psi^{(1,0,1)}$  & $\psi^{(1,2,1)}$ & $\psi^{(1,0,1)}$  \\
\hline
 1 &       413.1586 &         8.8245 &      -289.4977 &         1.5578 &       244.8123 &         3.7011 \\
 2 &     -2384.9885 &      -114.1754 &      1734.9402 &       -22.5966 &     -1545.0640 &       -63.3077 \\
 3 &      4895.6915 &       524.1161 &     -3619.8761 &       127.1963 &      3437.6769 &       378.8322 \\
 4 &     -4120.8799 &     -1159.3614 &      2815.1704 &      -371.5994 &     -2942.2393 &     -1087.4587 \\
 5 &       490.3119 &      1390.2561 &       415.7821 &       619.5920 &      -231.3078 &      1690.6255 \\
 6 &      1523.0854 &      -947.6269 &     -2036.6024 &      -602.2710 &      2146.9545 &     -1488.8664 \\
 7 &     -1076.5260 &       366.2696 &      1288.9620 &       332.9247 &     -1496.2120 &       736.8267 \\
 8 &       293.9187 &       -74.6692 &      -343.3866 &       -96.0393 &       431.2140 &      -189.8749 \\
 9 &       -29.5824 &         6.2449 &        34.1359 &        11.1873 &       -46.2070 &        19.7831 \\
\hline
\end{tabular}
}
\end{center}
\end{table}

\begin{table}
\caption{Constants $c_i(S,L,J;n)$ from Eq. (\ref{a-1}) (in GeV)
for the wave functions of $\chi_{c2}$-mesons in solution $I(c\bar c)$}
\begin{center} {\scriptsize
\begin{tabular}{|r|r|r|r|r|r|r|}
\hline
& \multicolumn{2}{|c|}{$\chi_{c2}(1P)$} & \multicolumn{2}{|c|}{$\chi_{c2}(2P)$} &
  \multicolumn{2}{|c|}{$\chi_{c2}(3P)$} \\
\hline
 & \multicolumn{2}{|c|}{$W_{11}\qquad W_{13}\qquad W_{33}$} &
   \multicolumn{2}{|c|}{$W_{11}\qquad W_{13}\qquad W_{33}$} &
   \multicolumn{2}{|c|}{$W_{11}\qquad W_{13}\qquad W_{33}$} \\
 & \multicolumn{2}{|c|}{0.99952 \, -0.00070 \, 0.00118} &
   \multicolumn{2}{|c|}{0.99765 \, -0.00211 \, 0.00446} &
   \multicolumn{2}{|c|}{0.99209 \, -0.00518 \, 0.01309} \\
\hline
$i$ & $\psi^{(1,1,2)}$ & $\psi^{(1,3,2)}$ & $\psi^{(1,1,2)}$ & $\psi^{(1,3,2)}$  & $\psi^{(1,1,2)}$ & $\psi^{(1,3,2)}$  \\
\hline
 1 &       -17.5035 &         9.0856 &         4.5957 &        39.0522 &       -93.1247 &        44.4200 \\
 2 &       -86.9407 &       -40.9452 &      -623.9524 &      -250.9432 &       163.8698 &      -311.5971 \\
 3 &       506.5360 &        85.1047 &      3316.2121 &       629.3054 &      1317.7565 &       796.6134 \\
 4 &      -988.9820 &      -100.6670 &     -7222.9252 &      -821.2865 &     -5066.0027 &      -990.8076 \\
 5 &      1026.3189 &        71.0253 &      8391.4505 &       608.6930 &      7472.6180 &       644.8615 \\
 6 &      -625.3392 &       -29.4658 &     -5639.3954 &      -257.5492 &     -5744.7189 &      -202.3282 \\
 7 &       224.1992 &         6.6484 &      2197.9985 &        57.7459 &      2434.8192 &        13.3976 \\
 8 &       -43.7175 &        -0.6473 &      -460.8963 &        -5.2652 &      -539.0269 &         7.5182 \\
 9 &         3.5664 &         0.0062 &        40.1760 &        -0.0143 &        48.6569 &        -1.3447 \\
\hline
\hline
& \multicolumn{2}{|c|}{$\chi_{c2}(4P)$} & \multicolumn{2}{|c|}{$\chi_{c2}(5P)$} &
  \multicolumn{2}{|c|}{$\chi_{c2}(6P)$} \\
\hline
 & \multicolumn{2}{|c|}{$W_{11}\qquad W_{13}\qquad W_{33}$} &
   \multicolumn{2}{|c|}{$W_{11}\qquad W_{13}\qquad W_{33}$} &
   \multicolumn{2}{|c|}{$W_{11}\qquad W_{13}\qquad W_{33}$} \\
 & \multicolumn{2}{|c|}{0.88640 \, -0.00477 \, 0.11837} &
   \multicolumn{2}{|c|}{0.98493 \, -0.01273 \, 0.02780} &
   \multicolumn{2}{|c|}{0.93818 \, -0.03946 \, 0.10128} \\
\hline
$i$ & $\psi^{(1,1,2)}$ & $\psi^{(1,3,2)}$ & $\psi^{(1,1,2)}$ & $\psi^{(1,3,2)}$  & $\psi^{(1,1,2)}$ & $\psi^{(1,3,2)}$  \\
\hline
 1 &       638.6609 &       391.6378 &      -134.7570 &      -221.6602 &       117.7874 &       104.0293 \\
 2 &     -5864.8034 &     -2008.6939 &      1751.9157 &      1250.1138 &     -1488.3254 &      -706.9778 \\
 3 &     21108.8607 &      4224.3629 &     -8121.0456 &     -2766.6069 &      6956.8543 &      1877.3877 \\
 4 &    -39600.9145 &     -4709.0719 &     18375.8266 &      3074.1617 &    -16174.3244 &     -2509.6472 \\
 5 &     42944.0995 &      2957.4768 &    -22896.3579 &     -1771.0161 &     20877.3955 &      1793.4813 \\
 6 &    -27861.8978 &      -999.8571 &     16466.2322 &       434.8957 &    -15563.0676 &      -647.3392 \\
 7 &     10649.1787 &       137.5405 &     -6796.9738 &        31.0544 &      6631.3660 &        78.8171 \\
 8 &     -2206.5653 &         8.3919 &      1492.7938 &       -36.5787 &     -1494.2661 &        13.2404 \\
 9 &       190.7603 &        -3.1287 &      -134.9738 &         5.1157 &       137.6804 &        -3.2854 \\
\hline
\hline
& \multicolumn{2}{|c|}{$\chi_{c2}(1F)$} & \multicolumn{2}{|c|}{$\chi_{c2}(2F)$} &
  \multicolumn{2}{|c|}{$\chi_{c2}(3F)$} \\
\hline
 & \multicolumn{2}{|c|}{$W_{11}\qquad W_{13}\qquad W_{33}$} &
   \multicolumn{2}{|c|}{$W_{11}\qquad W_{13}\qquad W_{33}$} &
   \multicolumn{2}{|c|}{$W_{11}\qquad W_{13}\qquad W_{33}$} \\
 & \multicolumn{2}{|c|}{0.00301 \, -0.00135 \, 0.99834} &
   \multicolumn{2}{|c|}{0.10992 \, -0.00454 \, 0.89462} &
   \multicolumn{2}{|c|}{0.03903 \, -0.01166 \, 0.97263} \\
\hline
$i$ & $\psi^{(1,3,2)}$ & $\psi^{(1,1,2)}$ & $\psi^{(1,3,2)}$ & $\psi^{(1,1,2)}$  & $\psi^{(1,3,2)}$ & $\psi^{(1,1,2)}$  \\
\hline
 1 &      -242.7471 &        10.5652 &      -565.8475 &       225.5515 &       733.1238 &        19.8324 \\
 2 &       744.2459 &       -95.8214 &      2117.9502 &     -2067.2776 &     -3370.3257 &      -120.2819 \\
 3 &      -845.6564 &       355.4159 &     -2625.9078 &      7468.3409 &      5512.4695 &       179.6152 \\
 4 &       320.2998 &      -686.5646 &       638.7903 &    -14099.5057 &     -3448.4647 &       157.0118 \\
 5 &       166.8886 &       763.7615 &      1409.3867 &     15396.8275 &      -410.2084 &      -688.2629 \\
 6 &      -224.3091 &      -506.8350 &     -1486.9023 &    -10057.8914 &      1736.5500 &       755.6477 \\
 7 &        96.3257 &       197.7245 &       638.6645 &      3868.9470 &      -956.7876 &      -394.8955 \\
 8 &       -19.3136 &       -41.7533 &      -131.7223 &      -806.4039 &       226.5196 &       101.4201 \\
 9 &         1.5203 &         3.6750 &        10.7017 &        70.0913 &       -20.3853 &       -10.2876 \\
\hline
\hline
& \multicolumn{2}{|c|}{$\chi_{c2}(4F)$} & \multicolumn{2}{|c|}{$\chi_{c2}(5F)$} &
  \multicolumn{2}{|c|}{$\chi_{c2}(6F)$} \\
\hline
 & \multicolumn{2}{|c|}{$W_{11}\qquad W_{13}\qquad W_{33}$} &
   \multicolumn{2}{|c|}{$W_{11}\qquad W_{13}\qquad W_{33}$} &
   \multicolumn{2}{|c|}{$W_{11}\qquad W_{13}\qquad W_{33}$} \\
 & \multicolumn{2}{|c|}{0.01125 \, -0.00896 \, 0.99771} &
   \multicolumn{2}{|c|}{0.01670 \, -0.01856 \, 1.00186} &
   \multicolumn{2}{|c|}{0.01548 \, -0.02033 \, 1.00485} \\
\hline
$i$ & $\psi^{(1,3,2)}$ & $\psi^{(1,1,2)}$ & $\psi^{(1,3,2)}$ & $\psi^{(1,1,2)}$  & $\psi^{(1,3,2)}$ & $\psi^{(1,1,2)}$  \\
\hline
 1 &      -580.3877 &       -12.4060 &       431.6178 &         1.5693 &       345.9078 &         5.2688 \\
 2 &      2948.0418 &       137.4061 &     -2289.0231 &       -16.1472 &     -1924.9704 &       -74.4252 \\
 3 &     -5368.6253 &      -566.3875 &      4363.4753 &        50.9676 &      3911.9978 &       376.9102 \\
 4 &      3893.8813 &      1161.4946 &     -3306.3734 &       -47.7121 &     -3311.2035 &      -935.6268 \\
 5 &         3.8221 &     -1320.6326 &       -27.4883 &       -46.6674 &       358.9932 &      1279.4826 \\
 6 &     -1709.0721 &       866.8103 &      1606.1715 &       127.5496 &      1343.8856 &     -1004.6984 \\
 7 &      1047.0953 &      -325.6228 &     -1012.8337 &       -98.5000 &      -964.7262 &       448.1251 \\
 8 &      -263.5517 &        64.7935 &       261.2351 &        33.0133 &       268.3202 &      -104.9359 \\
 9 &        24.8409 &        -5.2798 &       -25.0732 &        -4.1196 &       -27.3899 &         9.9921 \\
\hline
\end{tabular}
}
\end{center}
\end{table}

\begin{table}
\caption{Constants $c_i(S,L,J;n)$ from eq. (\ref{a-1}) (in GeV) for
the wave functions of $\eta_{c0}$, $\chi_{c0}$, $\chi_{c1}$ and
$h_{c1}$ mesons in solution $I(c\bar c)$}
\begin{center}
{\footnotesize \begin{tabular}{|r|r|r|r|r|r|r|} \hline $i$ &
$\eta_{c0}(1P)$ & $\eta_{c0}(2P)$ & $\eta_{c0}(3P)$ &
      $\eta_{c0}(4P)$ & $\eta_{c0}(5P)$ & $\eta_{c0}(6P)$  \\
\hline
 1 &         7.6494 &       -10.5636 &        18.1089 &        33.6294 &       -89.2312 &       -75.5014  \\
 2 &       -23.0211 &       -53.9217 &        15.0118 &      -170.0167 &      1032.3991 &      1029.1523  \\
 3 &       124.2754 &       461.0285 &      -623.7988 &      -168.7223 &     -4276.2522 &     -5025.8538  \\
 4 &      -395.6413 &     -1133.8010 &      1908.5781 &      2060.7362 &      8662.4563 &     11959.9547  \\
 5 &       656.5275 &      1456.7298 &     -2536.9289 &     -4211.5707 &     -9688.9143 &    -15569.2648  \\
 6 &      -605.5583 &     -1101.1359 &      1780.5781 &      3997.5023 &      6289.9521 &     11572.6965  \\
 7 &       314.7296 &       493.4357 &      -680.8186 &     -1997.5324 &     -2365.3552 &     -4872.6646  \\
 8 &       -86.2318 &      -121.2544 &       132.0261 &       509.0043 &       479.4585 &      1076.4010  \\
 9 &         9.7057 &        12.6004 &        -9.8555 &       -52.2319 &       -40.7532 &       -96.3877  \\
\hline
\hline
$i$ & $\chi_{c0}(1P)$ & $\chi_{c0}(2P)$ & $\chi_{c0}(3P)$ &
      $\chi_{c0}(4P)$ & $\chi_{c0}(5P)$ & $\chi_{c0}(6P)$  \\
\hline
 1 &       -17.8495 &       -50.2837 &      -151.3174 &      -313.0348 &       318.7744 &      -249.2578  \\
 2 &         3.8696 &         5.7940 &       731.5190 &      2334.3640 &     -2879.7647 &      2448.4531  \\
 3 &        29.1978 &       634.0603 &     -1074.8942 &     -6487.9099 &      9875.5585 &     -9126.1510  \\
 4 &        70.7786 &     -1637.8698 &       270.8311 &      8844.2446 &    -17101.9307 &     17108.1285  \\
 5 &      -273.1091 &      1949.8183 &       727.7711 &     -6388.0162 &     16610.2166 &    -17802.3919  \\
 6 &       321.4758 &     -1315.0808 &      -767.5997 &      2335.4592 &     -9396.2245 &     10577.7296  \\
 7 &      -184.1650 &       521.0477 &       313.2245 &      -289.9459 &      3061.3295 &     -3502.4000  \\
 8 &        52.5331 &      -113.9435 &       -55.4903 &       -50.4060 &      -530.9090 &       584.2914  \\
 9 &        -5.9970 &        10.6960 &         3.0610 &        13.0531 &        37.9718 &       -36.1088  \\
\hline
\hline
$i$ & $\chi_{c1}(1P)$ & $\chi_{c1}(2P)$ & $\chi_{c1}(3P)$ &
      $\chi_{c1}(4P)$ & $\chi_{c1}(5P)$ & $\chi_{c1}(6P)$  \\
\hline
 1 &        23.0070 &        56.7513 &       217.3310 &      -388.4208 &       344.6565 &       238.1913  \\
 2 &        -3.8670 &       -14.0364 &     -1332.2767 &      3249.2463 &     -3365.3754 &     -2490.3066  \\
 3 &      -115.4587 &      -778.7437 &      3058.3410 &    -10453.2199 &     12574.5044 &      9950.4054  \\
 4 &       191.1470 &      2115.0575 &     -3497.5882 &     17295.2284 &    -23976.3393 &    -20229.2674  \\
 5 &      -112.1101 &     -2567.5452 &      2172.1531 &    -16412.7705 &     25956.4509 &     23216.4628  \\
 6 &         4.4321 &      1719.1521 &      -716.3965 &      9284.6416 &    -16581.9431 &    -15583.6367  \\
 7 &        23.5865 &      -657.5312 &       101.6710 &     -3089.5089 &      6182.0939 &      6036.8548  \\
 8 &        -9.8945 &       134.9058 &         1.4698 &       556.8344 &     -1242.3831 &     -1245.5237  \\
 9 &         1.2824 &       -11.5449 &        -1.2720 &       -41.8049 &       103.9046 &       105.5852  \\
\hline
\hline
$i$ & $h_c(1P)$ & $h_c(2P)$ & $h_c(3P)$ &
      $h_c(4P)$ & $h_c(5P)$ & $h_c(6P)$  \\
\hline
 1 &       -19.4824 &        44.2886 &       130.8468  &      -296.9419 &       431.5082 &      -293.9990  \\
 2 &       -42.1580 &       119.7469 &      -553.3042  &      2284.4817 &     -4002.0767 &      2983.8574  \\
 3 &       305.3320 &     -1319.3114 &       298.1904  &     -6661.4528 &     14363.1096 &    -11699.4557  \\
 4 &      -573.6405 &      3202.4192 &      1602.7564  &      9794.3567 &    -26533.5962 &     23525.9355  \\
 5 &       548.0217 &     -3792.3619 &     -3269.2777  &     -8048.5695 &     28014.7411 &    -26866.9716  \\
 6 &      -299.4493 &      2527.4497 &      2752.3010  &      3805.6849 &    -17540.3649 &     18029.3913  \\
 7 &        93.8170 &      -967.1883 &     -1199.8706  &     -1003.0218 &      6432.1760 &     -7008.3077  \\
 8 &       -15.4547 &       198.4952 &       266.1679  &       130.1848 &     -1274.7649 &      1455.2756  \\
 9 &         1.0081 &       -16.9394 &       -23.7422  &        -5.6569 &       105.3324 &      -124.4976  \\
\hline
\end{tabular}
}
\end{center}
\end{table}

\begin{figure}
\centerline{\epsfig{file=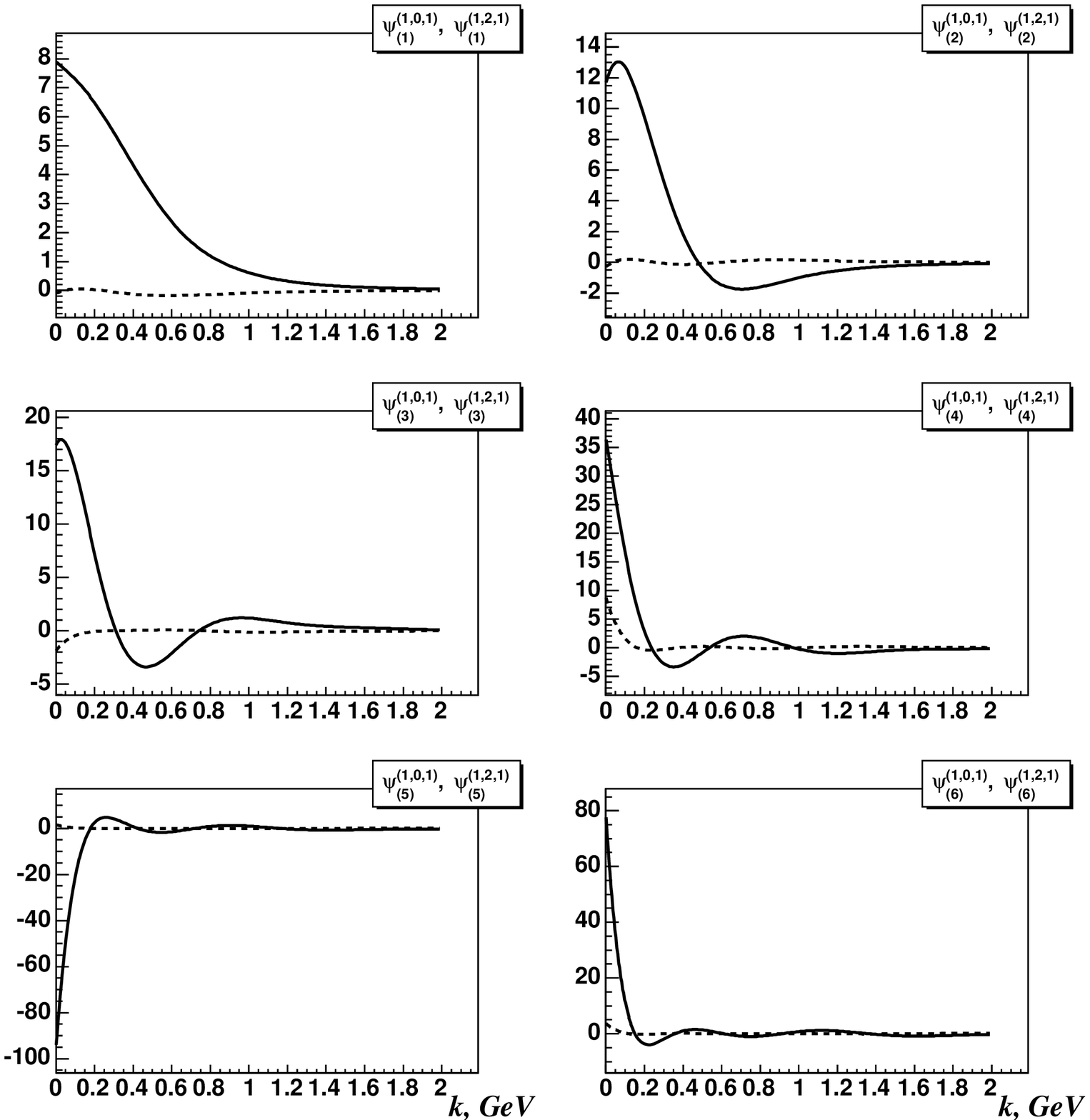,width=13cm}}
\caption{Wave functions for $\psi(nS)$ in solution $ I(c\bar c)$.
Solid lines stand for $\psi^{(1,0,1)}_n$ and dashed ones for
$\psi^{(1,2,1)}_n$. All magnitudes are in GeV units. }
\end{figure}

\begin{figure}
\centerline{\epsfig{file=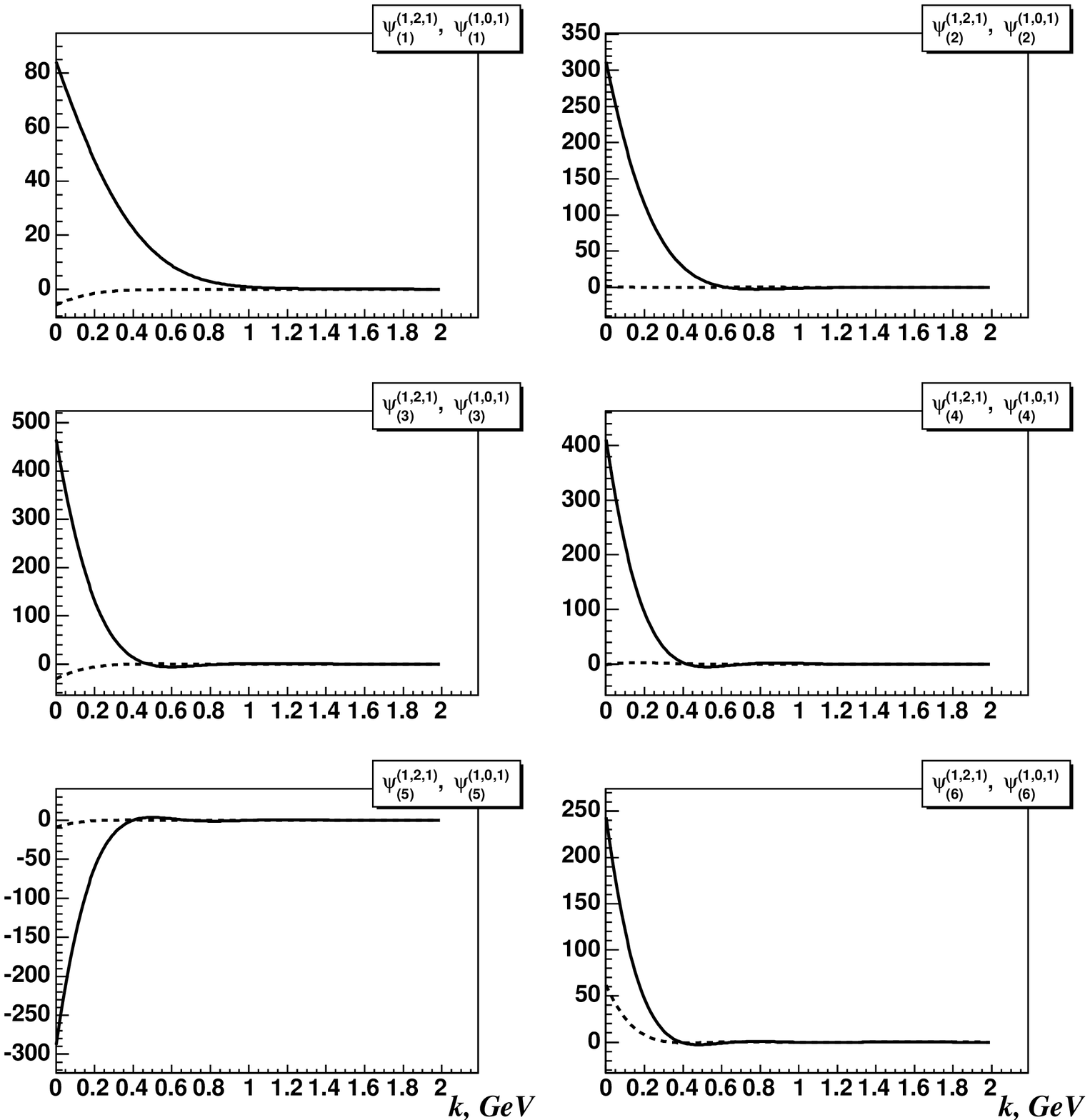,width=13cm}}
\caption{Wave functions for $\psi(nD)$ in solution $ I(c\bar c)$.
Solid lines stand for $\psi^{(1,2,1)}_n$ and dashed ones for
$\psi^{(1,0,1)}_n$. All magnitudes are in GeV units.}
 \end{figure}

\begin{figure}
\centerline{\epsfig{file=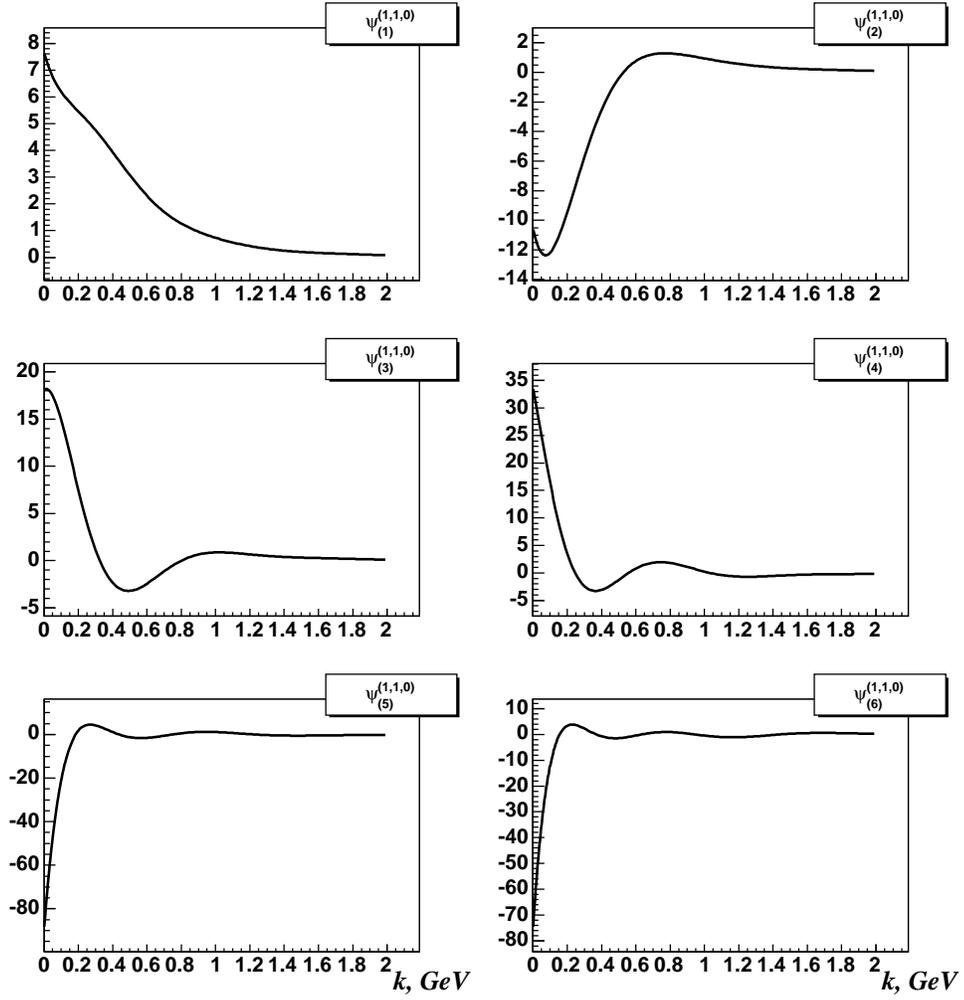,width=13cm}}
\caption{Wave functions for $\eta_c$ in solution $I(c\bar c)$.}
\end{figure}

\begin{figure}
\centerline{\epsfig{file=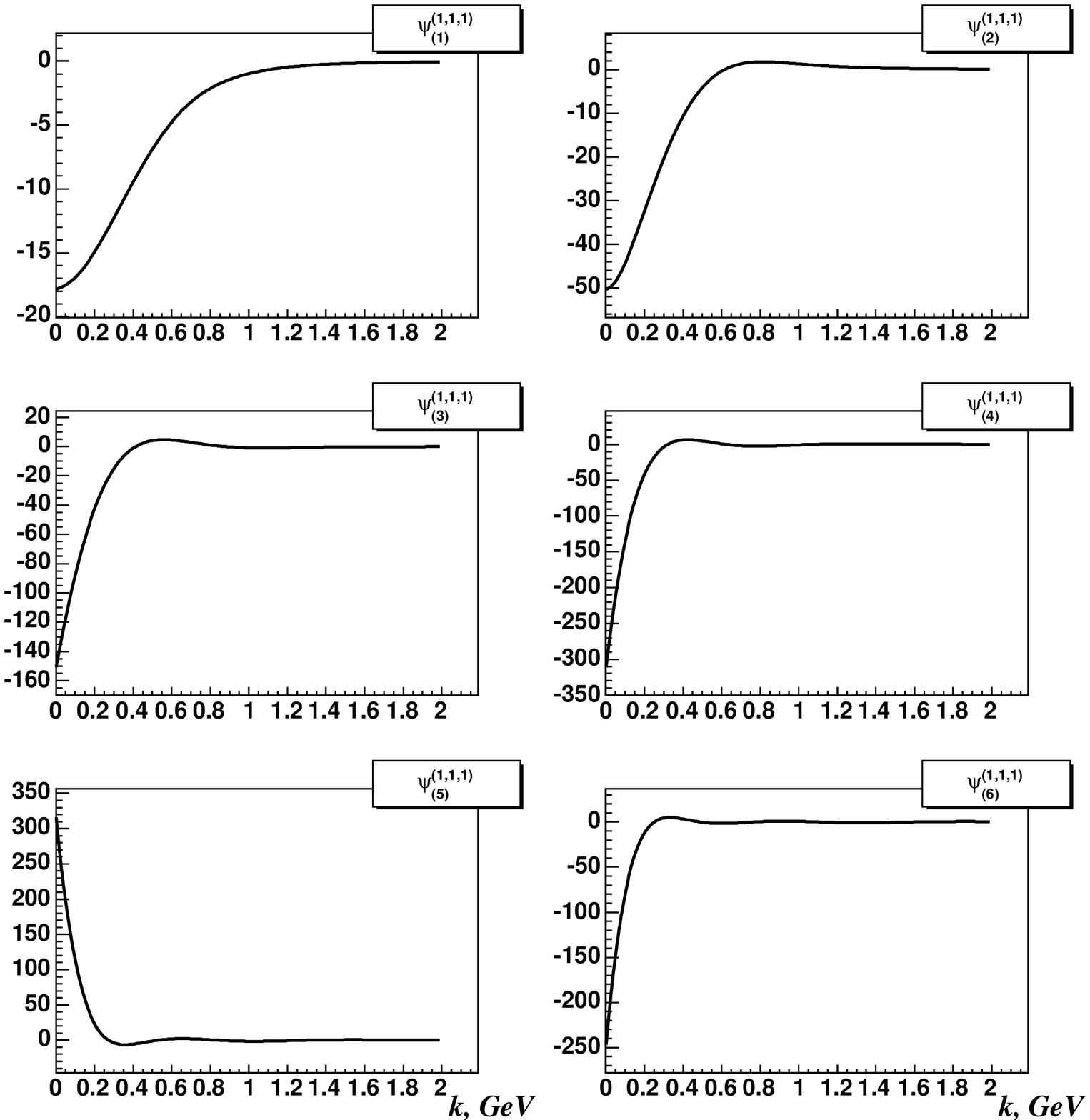,width=13cm}}
\caption{Wave functions for $\chi_{c0}$ in solution $I(c\bar c)$.}
\end{figure}

\begin{figure}
\centerline{\epsfig{file=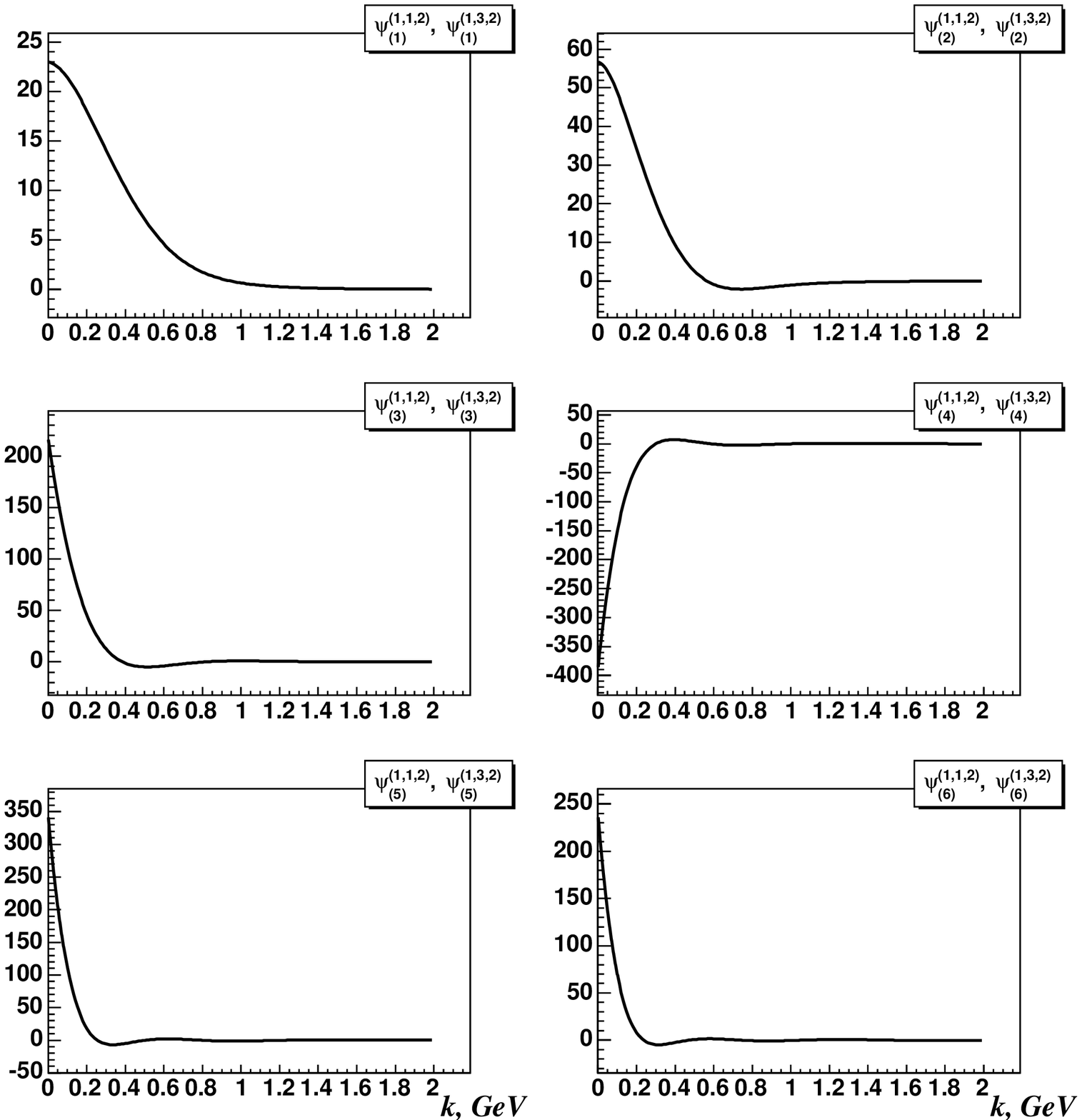,width=13cm}}
\caption{Wave functions for $\chi_{c1}$ in solution $I(c\bar c)$.}
\end{figure}

\begin{figure}
\centerline{\epsfig{file=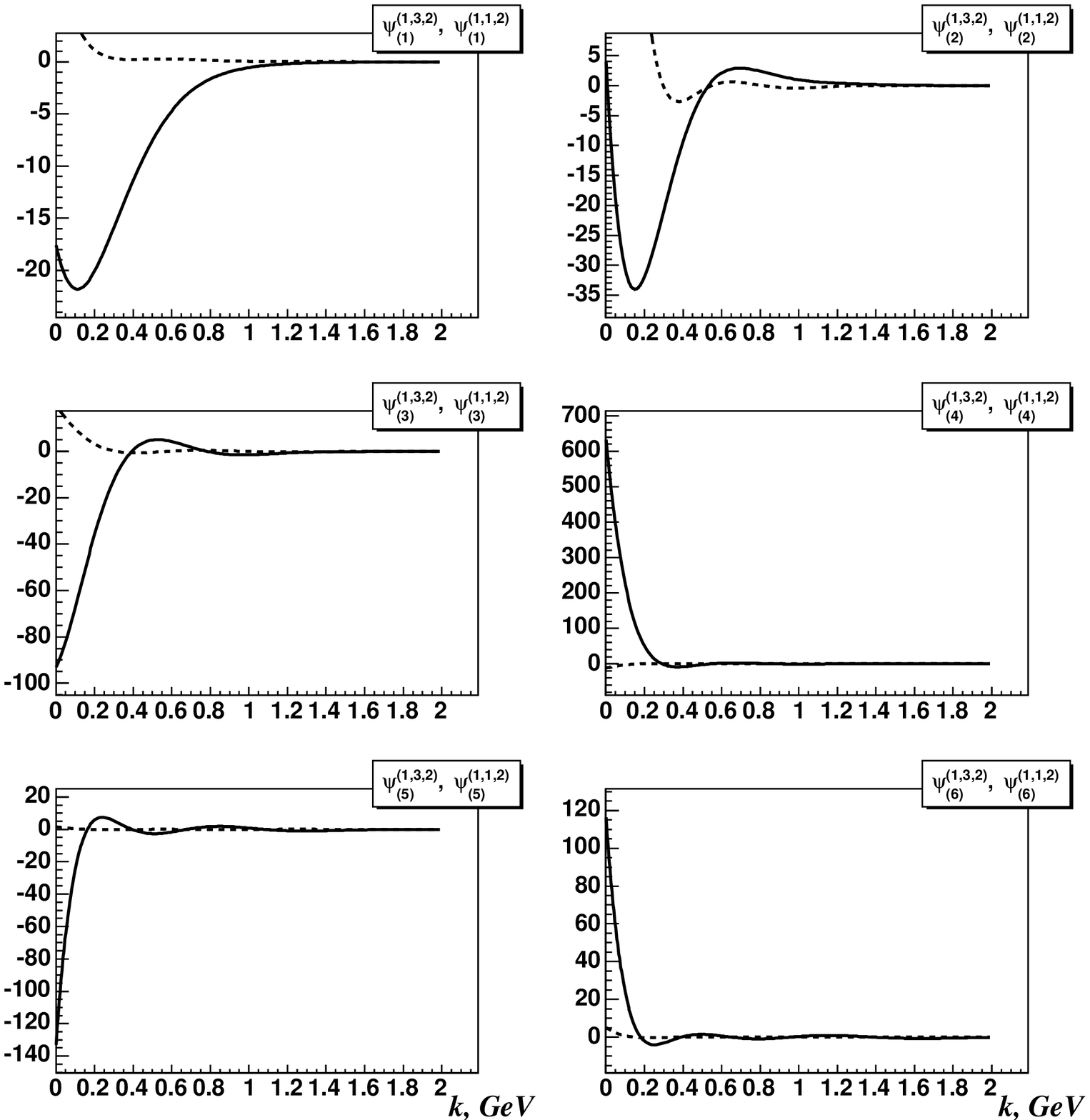,width=13cm}}
\caption{Wave functions for $\chi_{c2}$ in solution $I(c\bar c)$.
Solid lines stand for $\psi^{(1,1,2)}_n$ and dashed ones for
$\psi^{(1,3,2)}_n$. }
\end{figure}

\begin{figure}
\centerline{\epsfig{file=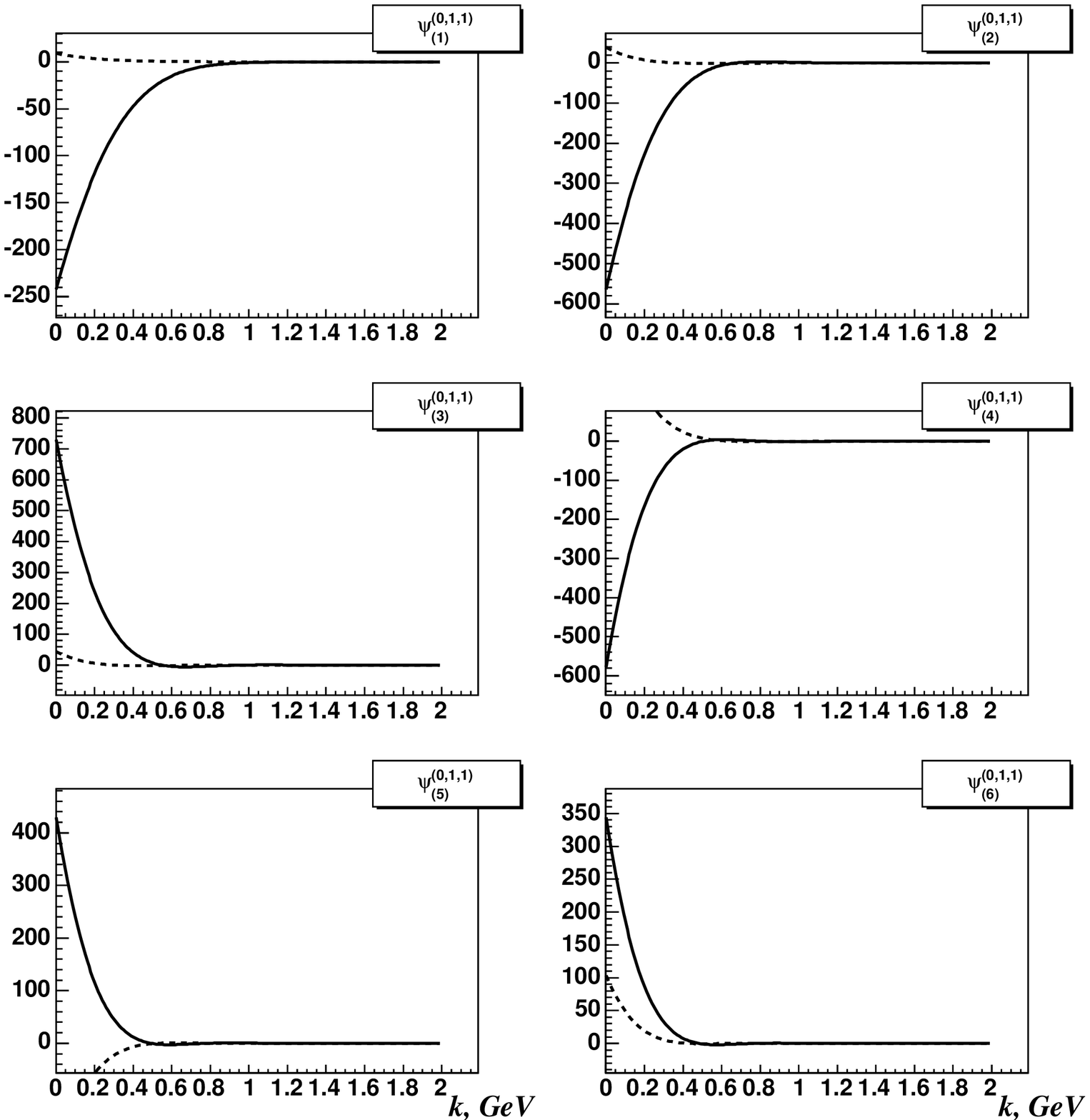,width=13cm}}
\caption{Wave functions for $\chi_{c2}$ in Solution $I(c\bar c)$.
Solid lines stand for $\psi^{(1,3,2)}_n$ and dashed ones for
$\psi^{(1,1,2)}_n$. }
\end{figure}


\clearpage
\begin{thebibliography}{99}
 \fussy
\bibitem{bb} V.V. Anisovich, L.G. Dakhno, M.A. Matveev,
 V.A. Nikonov, and  A.V. Sarantsev, hep-ph/0510410.
\bibitem{BS}
 A.V. Anisovich, V.V. Anisovich, B.N. Markov,  M.A. Matveev, and
A.V. Sarantsev,
 Yad. Fiz. {\bf 67}, 794 (2004)
[Phys. of Atomic Nuclei, {\bf 67}, 773  (2004)].

\bibitem{deut} V.V. Anisovich, M.N. Kobrinsky, D.I. Melikhov, and
A.V. Sarantsev, Nucl. Phys. A {\bf 544}, 747 (1992);\\
A.V. Anisovich and V.A. Sadovnikova, Yad. Fiz.
{\bf 55}, 2657 (1992); {\bf 57}, 75 (1994); Eur. Phys. J. A {\bf 2},
199 (1998).

\bibitem{physrev} V.V. Anisovich, D.I. Melikhov, and V.A. Nikonov,
Phys. Rev.  D {\bf52}, 5295 (1995); D {\bf55}, 2918 (1997).

\bibitem{epja} A.V. Anisovich, V.V. Anisovich, and V.A. Nikonov,
 Eur. Phys. J. A {\bf 12}, 103 (2001).

\bibitem{YFscalar} A.V. Anisovich, V.V. Anisovich, V.N. Markov,
and V.A. Nikonov,
Yad. Fiz. {\bf 65}, 523 (2002) [Phys. Atom. Nucl.
{\bf 65}, 497 (2002)].

\bibitem{YFtensor}
A.V. Anisovich, V.V. Anisovich, M.A. Matveev, and V.A. Nikonov,
Yad. Fiz. {\bf 66}, 946 (2003)
[Phys. Atom. Nucl.
{\bf 66}, 914 (2003)].

\bibitem{YF-g}
V.V. Anisovich, L.G. Dakhno, M.N. Markov, V.A. Nikonov, and
A.V. Sarantsev, Yad. Fiz., in press; hep-ph/0406320,
hep-ph/0410361.

\bibitem{Choi} S.K. Choi {\it at. al.}, (Belle Collab.), Phys. Rev.
Lett. {\bf 91}, 262001 (2003).

\bibitem{Acosta} D. Acosta {\it at. al.}, (CDF II Collab.), Phys. Rev.
Lett. {\bf 93}, 072001 (2004).


\bibitem{Abazov} V.M.Abrazov {\it at. al.}, (D0 Collab.), Phys. Rev.
Lett. {\bf 93}, 162002 (2004).

\bibitem{Aubert} B. Aubert {\it at. al.}, (Barbar Collab.), Phys. Rev.
D {\bf 71}, 071103 (2005).



\bibitem {5} N.A. T\" ornqvist, arXiv:hep-ph/0402237.

\bibitem {6} F.E. Close and P.R. Page, Phys. Lett. B {\bf 578}, 119
(2004).

\bibitem {7} M.B. Voloshin, Phys. Lett.
B {\bf 579}, 316 (2004).

\bibitem {8} C.Z. Yuan, X.M. Mo and P. Wang, Phys. Lett. B {\bf 579},
74 (2004).

\bibitem {9} C.-Y. Wong, Phys. Rev. C {\bf 69}, 055202 (2004).

\bibitem {10} T. Barnes and S. Godfrey, Phys. Rev. D {\bf 69}, 054008
(2004).

\bibitem {11} E.S. Swanson, Phys. Lett. B {\bf 588}, 189 (2004).

\bibitem {12} T. Skwarnicki, arXiv:hep-ph/0311243.

\bibitem {13} P. Bicudo, arXiv:hep-ph/0401106.

\bibitem {14} E.J. Eichten, K. Lane and C. Quigg, Phys. Rev. D
{\bf 69}, 094019 (2004).
\bibitem{Isgur} N. Isgur and M.B. Wiss, Phys. Lett. B {\bf 232}, 113
(1989); Phys. Lett. B {\bf 237}, 527 (1990).

\bibitem{Monohar} A.V. Monohar and C.T. Sachrajda, Phys. Lett. B
{\bf 592}, 473 (2004).

\bibitem{PDG}S. Eidelman, {\it et al.,} Phys. Lett. B {\bf 592}, 1 (2004).

\bibitem{Bugg} D.V. Bugg, Phys. Rev. D {\bf 71}, 016006, (2005).

\bibitem{Barnes} T. Barnes and S. Godfrey,  Phys. Rev. D {\bf 69},
054008, (2004).
\bibitem{raddecay}
 A.V. Anisovich, V.V. Anisovich, M.A. Matveev, V.N. Markov,
 V.A. Nikonov, and  A.V. Sarantsev, J. Phys. G: Nucl. Part. Phys., in
 press; hep-ph/0509042.


\bibitem{c1-1} J. Gaiser, {\it et al.,}  Phys. Rev. D {\bf 34}, 711 (1986).

\bibitem{c1-2} C.J. Biddick, {\it et al.,} Phys. Rev. Lett. {\bf 38},
1324 (1977).
\bibitem{Hernandez} J.J. Hern\'andez-Rey, S. Navas, and C. Patrignani,
Phys. Lett. B {\bf 952}, 822 (2004).



\bibitem{L3}  M. Acciari,  {\it et al.,} Phys. Lett. B {\bf 453},  73
(1999).

\bibitem{OPAL}K. Ackerstaff,  {\it et al.,} Phys. Lett. B {\bf 439}, 197
(1998).

\bibitem{CLEO}J. Dominick,  {\it et al.,} Phys. Rev. D {\bf 50}, 4265
(1994).

\bibitem{E760}T.A. Armstrong,  {\it et al.,} Phys. Rev. Lett. {\bf 70},
 2988 (1993).


\bibitem{Linde} J. Linde and H. Snellman,
Nucl. Phys. A {\bf 619}, 346 (1997).

\bibitem{Resag} J. Resag and C.R. M\"unz, Nucl. Phys. A {\bf 590}, 735
(1995).

\bibitem{NR} M.Beyer, U. Bohn, M.G. Huber, B.C. Metsch, and J. Resag,
Z.Phys C {\bf 55}, 307 (1992).

\bibitem{deWitt} M.A. DeWitt, H.M. Choi and C.R. Ji,
Phys. Rev. D{\bf 68}: 054026 (2003).

\bibitem{Xiao} B.-W. Xiao and B.-Q. Ma, Phys. Rev. D {\bf 68}, 034020
(2003).

\bibitem{G}D. Ebert, R.N. Faustov, and V.O. Galkin,
Phys. Rev D {\bf 67}, 014027  (2003).

\bibitem{M}S.N. M\"unz, Nucl. Rhys. A {\bf 609}, 364 (1996).

\bibitem{Gupta}S.N. Gupta, S.F. Radford, and W.W. Repko, Phys. Rev.
 D {\bf 54}, 2075 (1996).

\bibitem{Sch}G.A. Schuler, F.A Berends, and R. van Gulik, Nucl. Rhys.
B {\bf 523}, 423 (1998).

\bibitem{Huang}H.-W. Huang, {\it et. al.} Phys. Rev.
D {\bf 54}, 2123 (1996); D {\bf 56}, 368 (1997).

\bibitem{Barnes2} E.S. Ackleh, T. Barnes, {\it et al.,} Phys. Rev.
D {\bf 45}, 232 (1992).


\end{thebibliography}
\end{document}